\documentclass[11pt]{article}
\usepackage{amsmath,amssymb,color,epsfig,cite}
\usepackage{xcolor}
\usepackage{mathrsfs}

\usepackage{mathtools}
\allowdisplaybreaks

\usepackage{amsfonts}
\def\td{\tilde}
\newcommand{\hoch}[1]{$\, ^{#1}$}

\makeatletter
\@addtoreset{equation}{section}
\makeatother

\newcommand{\be}{\begin{equation}}
\newcommand{\ee}{\end{equation}}
\newcommand{\bea} {\begin{eqnarray}}
\newcommand{\eea}{\end{eqnarray}}
\newcommand{\nn}{\nonumber}

\def\ft#1#2{{\textstyle{\frac{\scriptstyle #1}{\scriptstyle #2} } }}
\def\fft#1#2{{\frac{#1}{#2}}}
\def\dfft#1#2{{\displaystyle\fft{#1}{#2}}}

\def\0{{\sst{(0)}}}
\def\1{{\sst{(1)}}}
\def\2{{\sst{(2)}}}
\def\3{{\sst{(3)}}}
\def\4{{\sst{(4)}}}
\def\5{{\sst{(5)}}}
\def\6{{\sst{(6)}}}
\def\7{{\sst{(7)}}}
\def\8{{\sst{(8)}}}
\def\sst#1{{\scriptscriptstyle #1}}
\def\oneone{\rlap 1\mkern4mu{\rm l}}
\def\ep{{\epsilon}}
\def\del{{\partial}}

\def\u#1{{{\underline #1\,}}}

\def\cA{{{\cal A}}}

\def\cG{{{\cal G}}}
\def\cB{{{\cal B}}}
\def\cC{{{\cal C}}}

\def\im{{{\rm i\,}}}

\def\wtd{\widetilde}

\def\cE{{{\cal E}}}
\def\cO{{{\cal O}}}
\def\cx{{{\bf x}}}
\def\cL{{{\cal L}}}
\def\cS{{{\cal S}}}

\def\ie{{ i.e.~}}

\def\bmu{{{\bar\mu}}}
\def\bnu{{{\bar\nu}}}
\def\bphi{{{\bar\phi}}}

\def\bM{{{\bf M}}}

\def\rms{{{r_{\!\!\sst -}}}}
\def\rps{{{r_{\!\!\sst +}}}}

\def\ben{\begin{equation}}
\def\bea{\begin{eqnarray}}
\def\een{\end{equation}}
\def\eea{\end{eqnarray}}

\def \cL {{\cal  L}}
\def \cG {{\cal  G}}

\def\bvphi{{{\boldsymbol\varphi}}}

\def\ft#1#2{{\textstyle{\frac{\scriptstyle #1}{\scriptstyle #2} } }}
\def\fft#1#2{{\frac{#1}{#2}}}

\begin{document}

\begin{flushright}
\hfill {MI-HET-832
}\\
\end{flushright}

\begin{center}
{\large {\bf On The Perturbations of Gibbons-Maeda Black Holes
in Einstein-Maxwell-Dilaton Theories
 }}

\vspace{15pt}
{\large  C.N. Pope$^{1,2}$, D.O. Rohrer$^{1}$ and  B.F. Whiting$^{3}$}

\vspace{15pt}

\hoch{1}{\it George P. \& Cynthia Woods Mitchell  Institute
for Fundamental Physics and Astronomy,\\
Texas A\&M University, College Station, TX 77843, USA}

\hoch{2}{\it DAMTP, Centre for Mathematical Sciences,
 Cambridge University,\\  Wilberforce Road, Cambridge CB3 OWA, UK}

\hoch{3}{\it Department of Physics, P.O. Box 118440, University of Florida,\\
Gainesville, FL 32611-8440, USA}

\vspace{10pt}

%\vspace{30pt}

%\underline{ABSTRACT}

\end{center}

\begin{abstract}

{\normalsize

The study of perturbations around black hole backgrounds in general
relativity and Einstein-Maxwell theory 
has a long history, going back to the work of 
Regge and Wheeler in the 1950s. As part of a broader investigation of 
perturbations around black holes in supergravity, we describe here
our results for the perturbations around the Gibbons-Maeda static 
charged black holes in a class of Einstein-Maxwell-Dilaton theories.
Our analysis follows the general strategy developed by Chandrasekhar
and Xanthopoulos for the perturbations of the Reissner-Nordstr\"om 
black hole. Here, the analysis is considerably more involved, because
of the presence of the dilaton field, which couples to the other polar
modes.  We nonetheless find that the problem is completely solvable, in
the sense that one can separate variables and eventually describe all the
perturbations in terms of diagonalised second-order radial equations. 
We are able to prove the mode stability of all the Gibbons-Maeda 
black hole solutions.
}

\end{abstract}

{\small 
pope@physics.tamu.edu, rohrer@physics.tamu.edu, bernard@phys.ufl.edu }
\pagebreak

\tableofcontents
\addtocontents{toc}{\protect\setcounter{tocdepth}{2}}

\section{Introduction}

   The study of perturbations of  black hole solutions dates back to the
late 1950s, beginning with the seminal work of Regge and Wheeler 
\cite{regwhe}, who first studied the perturbations of the Schwarzschild
solution in general relativity.  (See also \cite{wheeler}, for an early
discussion of Maxwell's equations on a Schwarzschild background.) 
There was a great deal of activity
in the 1970s and 1980s, where the focus was on investigating the 
linearised perturbations of the Schwarzschild solution, 
of the charged Reissner-Nordstr\"om solution, and of the rotating 
Kerr solution. In all
three of these cases, a complete separation of variables and a decoupling
of the various modes of excitation can be achieved.  By contrast, this
has not been achieved for the case of the charged and rotating Kerr-Newman 
solution. 

    The black holes mentioned above are are solutions either of
pure Einstein gravity or, in the case of the charged black holes, the
Einstein-Maxwell theory.  In more recent times black hole solutions 
have also been found in supergravity theories, some of which arise
also as the low-energy effective limits of string theory or M-theory.  It
therefore becomes of interest to study also the perturbations of the
black hole solutions in these larger theories where, in particular, there
are scalar fields and, commonly, multiple gauge fields.  In higher
dimensions, there may be
other higher-form fields as well.  As is evident already from the situation for
the black holes in Einstein-Maxwell theory, things can quickly become
intractably complicated if one wants to consider solutions with both charge 
and angular momentum.  Thus the kinds of (charged) 
supergravity black holes for which one
can realistically hope to carry out a complete perturbation analysis
are likely to be restricted to those without rotation.   

    As a prelude to forthcoming work on black holes in supergravity theories,
in this paper we shall investigate a relatively modest extension
beyond the perturbations of the static Reissner-Nordstr\"om 
black hole of Einstein-Maxwell theory, by considering the static electrically
charged black holes in a class of four-dimensional 
Einstein-Maxwell-Dilaton (EMD) theories.
Although these EMD theories 
are not themselves supersymmetrisable, they do nevertheless
represent a substantial extension, beyond Einstein-Maxwell theory, which
introduces the additional intricacy of a scalar field that couples both
to gravity and to the Maxwell field, thus adding greatly to the inter-coupling
of the perturbative modes that need to be studied.  As such, the new features
that we encounter in the EMD analysis are likely to be indicative of
what one may expect to find also in the analysis for supergravity
black holes.
   
    As we shall show, similar techniques to those employed by 
Chandrasekhar, Xanthopoulos and Wald in a series of papers studying
the perturbations of Schwarzschild and Reissner-Nordstr\"om black holes
\cite{chanxant,xanth1,xanth2,wald,wald2} can be successfully applied to
the problem of separating variables and subsequently decoupling the
perturbative modes in the case of the static EMD black holes.  Interestingly,
the equations that arise in this case are of a considerably greater complexity 
than those in the Reissner-Nordstr\"om case, and yet we find that the
problem of separating variables and decoupling the modes 
is still a completely tractable one.  Having obtained the
decoupled equations describing the linearised perturbations we can then
study the question of mode stability of the EMD black hole solutions, and
we are able to show that they are indeed all mode stable.

  The Lagrangian describing the class of EMD theories that we shall
consider takes the form
%%%%%
\bea
{\cal L}= \sqrt{-g}\, \big( R -\ft12 (\del\phi)^2 - e^{a\phi}\, F^2\big)\,,
\label{emdlag}
\eea
%%%%%%
where $F^2=F^{\mu\nu}\,F_{\mu\nu}$ and $a$ is the dilaton coupling 
constant.  This is a non-trivial parameter that characterises inequivalent
EMD theories, and its value can be freely chosen.  (Without loss of
generality, it may be taken to be non-negative.)
The equations of motion following from the EMD Lagrangian (\ref{emdlag})
are
%%%%%
\bea
R_{\mu\nu} &=& \ft12 \del_\mu\phi\, \del_\nu\phi + 2 e^{a\phi}\,
   (F_{\mu\rho}\, F_\nu{}^\rho - \ft14 F^{\rho\sigma}\, F_{\rho\sigma}\,
g_{\mu\nu})\,,\nn\\
\nabla_\nu (e^{a\phi}\, F^{\mu\nu}) &=&0\,,\nn\\
\square\phi &=& a e^{a\phi}\, F^{\mu\nu}\,F_{\mu\nu}\,.\label{emdeom}
\eea
%%%%%
If we take $a=0$ the system reduces to Einstein-Maxwell theory together
with a massless scalar field that couples only to gravity.

  The static electrically-charged black hole solutions of the EMD theory
(\ref{emdlag}) were constructed by Gibbons and Maeda \cite{gibmae}.  In
the notation we shall be using here, these solutions were presented in
\cite{cvegibpop} in the form:\footnote{The solutions appear in
section 3.3.2 of \cite{cvegibpop}. Note that $\phi$ in that
paper is equal to $\ft12$ of the $\phi$ we are using here.  The solution
in \cite{cvegibpop} is given for a magnetically-charged black
hole.  A straightforward dualisation converts it to the electrically-charged
solution we are considering here.}
%%%%%%
\bea
ds^2 &=& -\Delta\, dt^2 + \fft{dr^2}{\Delta} + R^2\, (d\theta^2 +
\sin^2\theta\, d\varphi^2)\,,\nn\\
e^{-a\phi} &=& f_-^{\ft{2a^2}{1+a^2}}\,,\qquad A= \fft{Q}{r}\, dt\,,\nn\\
\Delta &=& f_+\, f_-^{\ft{1-a^2}{1+a^2}}\,,\qquad 
R = r\, f_-^{\ft{a^2}{1+a^2}}\,,\qquad f_\pm = 1-\fft{r_\pm}{r}\,.
\label{bhsol}
\eea
%%%%%
Note that $\Delta\, R^2$ is polynomial, as it is in Schwarzschild and 
Reissner-Nordstr\"om.
The electric charge $Q$ is related to the radii $r_\pm$ of the outer
and inner horizons by
%%%%%
\bea
Q^2= \fft{\rps\, \rms}{1+a^2}\,.
\eea
%%%%%
Note that in the limit when the dilaton coupling $a$ goes to zero, the
metric and Maxwell field reduce to those of the Reissner-Nordstr\"om
solution, and the dilaton field vanishes.

In the following sections, we shall set up the analysis of the linearised
perturbations around the Gibbons-Maeda black-hole solutions (\ref{bhsol}),
and then carry out the detailed process of separating the equations and
then disentangling the perturbations into decoupled sectors.  We shall 
follow the same general strategy that was employed by Chandrasekhar
and Xanthopoulos \cite{chanxant} when they studied the perturbations 
of the Reissner-Nordstr\"om black hole, in which they looked
directly at metric perturbations,
rather than using the Teukolsky-style approach of working with the 
gauge-invariant Newman-Penrose quantities. (As discussed in \cite{chandra},
the two approaches are closely related in these static black hole cases.)
As in the analysis in \cite{chanxant}, in our discussion of modes we
shall assume $\ell\ge2$.

As we shall show, the discussion can be separated into two disjoint
sectors, one of which is concerned with the axial fluctuation modes and
the other with the polar fluctuation modes.  (This also happened in
the simpler Reissner-Nordstr\"om case \cite{chanxant}.)  After
setting up the general formalism for describing the fluctuations around the
Gibbons-Maeda black holes, we then consider first the axial-mode sector,
which is the considerably simpler of the two sectors.  We show how the
equations describing the axial-mode fluctuations can be separated, and then
how the separated radial equations can be decoupled, giving rise to 
two diagonalised second-order equations.

    The analysis for the polar-mode sector is much more complicated. Here, we
show how the equations describing the polar fluctuations can be simplified
down to 
a system of seven coupled first-order differential equations.  This 
particular system of equations is what is known as {\it reducible}; a
procedure for analysing such systems was developed in various
papers, including \cite{wald,xanth1}, and it was applied in the case
of the Reissner-Nordstr\"om polar perturbations in \cite{xanth2}.  Here,
we carry out the procedure for the polar perturbations of the 
Gibbons-Maeda EMD black holes, showing how the seven first-order
equations can be solved in terms of the arbitrary solutions of three
second-order equations plus a seventh constant of integration characterising
a further particular solution of the first-order system.  We then show
how the three second-order equations can be diagonalised, thus arriving
at decoupled modes describing the polar perturbations.

  As we have mentioned, our analysis follows rather closely the
approach used in \cite{chanxant,xanth2}, and in fact some of our notation
is also close to that which was used in those papers.  However, when comparing
some of our expressions with those in \cite{chanxant,xanth2} it should
be borne in mind that we use a $(-+++)$ metric signature whereas they
used $(+---)$.  Also, our index numbering for the coordinates
$(t,r,\theta,\varphi)$ is $(0,1,2,3)$ whereas they used $(0,2,3,1)$.  
We also take care to indicate always whether numerical index values
refer to assignments for coordinate-index tensors or tetrad-index 
tensors.  Thus we denote the numerical components of a coordinate-index
vector $V_\mu$ by $(V_0,V_1,V_2,V_3)$, whereas the numerical
components of a tetrad-index vector $V_a$ are denoted by 
$(V_\u0,V_\u1,V_\u2,V_\u3)$.

\section{Linearised Perturbations} 

   An important simplification in the calculations 
can be made because the Gibbons-Maeda black hole background 
given in eqns (\ref{bhsol}) is spherically-symmetric.   This means that we 
can without loss of generality restrict attention to perturbations that are independent of the azimuthal angle $\varphi$.   As in the case of the
Reissner-Nordstr\"om background studied in \cite{chanxant,xanth2}, 
the perturbed metric
for discussing the fluctuations can therefore be taken to have the form
%%%%%
\bea
ds^2 = -e^{2\nu}\, dt^2 + e^{2\mu_1}\, dr^2 + e^{2\mu_2}\, d\theta^2 +
   e^{2\psi}\, (d\varphi - \cB_\alpha \, dx^\alpha)^2\,,\label{genmet}
\eea
%%%%%
where $\alpha=(0,1,2)$, and all quantities are taken to depend on
$t$, $r$ and $\theta$, but to be independent of
the azimuthal angle $\varphi$.  Note that this is very much in the style of
a Kaluza-Klein reduction ansatz from 4 to 3 dimensions, where the 
reduction is performed on the 
azimuthal coordinate $\varphi$.  (In \cite{chanxant,xanth2} 
the components $(\cB_0,\cB_1,\cB_2)$ of the
Kaluza-Klein vector $\cB_\alpha$ were named $(\omega,q_2,q_3)$ 
(with their index values 2 and 3 corresponding to our 1 and 2 ).)

As previously mentioned, the perturbations around the
background field configuration will decompose
into two disjoint sets, namely the {\it axial} perturbations,
and the {\it polar} perturbations.  These two classes of perturbation
are distinguished from one another by their behaviour under a reversal
of the sign of the azimuthal coordinate $\varphi$.  The equations for the
axial perturbations, which are of odd parity, 
decouple completely from the equations for the
polar perturbations, which are of even parity,
and so the two sets can be analysed independently.
As we already mentioned, the axial perturbations are much easier to 
deal with than the polar perturbations, and we shall discuss them first.  

  We shall choose the natural tetrad frame $e^a=e^a{}_\mu\, dx^\mu$ with
%%%%%
\bea
e^\u0= e^\nu\, dt\,,\qquad e^\u1= e^{\mu_1}\, dr \,,
  \qquad e^\u2= e^{\mu_2}\, d\theta\,,\qquad 
e^\u3= e^{\psi}\, (d\varphi -\cB_\alpha \, dx^\alpha)
\eea
%%%%%
in order to describe the metric (\ref{genmet}).
(One must be careful not to confuse the exponential functions
on the right-hand sides with the the tetrad symbols on the left-hand sides.)
Note that we are using the index $a$ to run over the tetrad-frame values
$(\u0,\u1,\u2,\u3)$ and the index $\mu$ to run over the coordinate-frame
values $(0,1,2,3)$.  The inverse tetrad $E_a{}^\mu$ can then be seen to be
given by
%%%%%
\bea
E_\u0{}^\mu\del_\mu &=& e^{-\nu}\,\Big(\fft{\del}{\del t} + \cB_0\,
\fft{\del}{\del\varphi}\Big)\,,\qquad
E_\u1{}^\mu\del_\mu= e^{-\mu_1}\, \Big(\fft{\del}{\del r} + \cB_1\,
\fft{\del}{\del\varphi}\Big)\,,\nn\\
E_\u2{}^\mu\del_\mu &=& e^{-\mu_2}\, \Big(\fft{\del}{\del \theta} + \cB_2\,
\fft{\del}{\del\varphi}\Big)\,,\qquad
E_\u3{}^\mu\, \del_\mu = e^{-\psi}\, \fft{\del}{\del\varphi}\,.
\eea
%%%%%
As can be verified,
%%%%%
\bea
e^a{}_\mu\, E_a{}^\nu=\delta_\mu^\nu\,,\qquad 
e^a{}_\mu\, E_b{}^\mu=\delta^a_b\,,\qquad
\eta_{ab}\, e^a{}_\mu\, e^b{}_\nu=g_{\mu\nu}\,,\qquad
g_{\mu\nu}\, E_a{}^\mu\, E_b{}^\nu=\eta_{ab}\,,
\eea
%%%%%
etc., where $\eta_{ab}=\hbox{diag}(-1,1,1,1)$ is the tetrad-frame 
Minkowski metric.

We now expand all the fields around their background values that they
take in the Gibbons-Maeda black holes, writing
%%%%%
\bea
\nu&=&\bar\nu + \ep\, \delta\nu\,,\qquad \mu_1=\bar\mu_1 + \ep\, \delta\mu_1
\,,\qquad \mu_2=\bar\mu_2 + \ep\, \delta \mu_2\,,\qquad
\psi=\bar\psi +\ep\, \delta\psi\,,\nn\\
\cB_\alpha &=& \ep\, \delta\cB_\alpha\,,\qquad
\phi =\bar\phi + \ep\, \delta\phi\,,\label{metricpert}
\eea
%%%%%
together with the field strength
%%%%%
\bea
F_{ab}= \bar F_{ab} + \ep\, \delta F_{ab}\,.\label{maxpert}
\eea
%%%%%
By comparing the these expressions, together with eqn (\ref{genmet}) 
with the black hole solution in eqns (\ref{bhsol}), we see that the
expressions for the background fields (denoted by the overbars) are
%%%%%
\bea
\bar\nu &=&\ft12\log\Delta\,,\qquad \bar\mu_1=-\bar\nu\,,\qquad
\bar\mu_2=\log R\,,\qquad \bar\psi=\log R+\log\sin\theta\,,\nn\\
\bar\cB_\alpha &=&0\,,\qquad
 a \bar\phi =2\log r-2\log R\,,\qquad \bar F_{\u0\u1} =\fft{Q}{r^2} 
\label{backgroundrels}
\eea
%%%%%
where $\Delta$ and $R$ are given in eqns (\ref{bhsol}), and all
components of the background field strength $\bar F_{ab}$ other than 
$ \bar F_{\u0\u1} =- \bar F_{\u1\u0}$ vanish. 

    In the linearised analysis that follows, we shall substitute the
perturbed expressions in eqns (\ref{metricpert}) and (\ref{maxpert})
into the field equations (\ref{emdeom}), keeping terms just
up to order $\epsilon$.  Following the strategy employed in 
\cite{chanxant,xanth2} for studying the Reissner-Nordstr\"om perturbations,
here too it is advantageous to write the EMD field equations using 
tetrad components, and expand these up to linear order in $\epsilon$.
Thus we substitute (\ref{metricpert}) and (\ref{maxpert}) into
the Einstein and Maxwell field equations, the Bianchi identities for
the Maxwell field, and the dilaton equation of motion, written in
the forms
%%%%%
\bea
(\hbox{Ein})_{ab} &\equiv& R_{ab} -\ft12 R\, \eta_{ab} 
-\ft12 \phi_a\, \phi_b +\ft14 \phi^c\,\phi_c\,\eta_{ab} \nn\\
&&-
      2 e^{a\phi}\, (F_{ac}\, F_b{}^c - \ft14 \eta_{ab} \, F^{cd} F_{cd})
  =0\,,
\label{tetradein}\\
(\hbox{Max})^a &\equiv& e^a{}_\mu\, \del_\nu(\sqrt{g}\, e^{a\phi}\,
        E^{b\mu}\, E^{c\nu}\, F_{bc})=0\,,\label{tetradmax}\\
(\hbox{Bianchi})^a &\equiv& e^a{}_\mu\, \varepsilon^{\mu\nu\rho\sigma}\,
\del_\nu(e^b{}_\rho\, e^c{}_\sigma\, F_{bc})=0\,,\label{tetradbianchi}\\
\square\phi &=& a\, e^{a\phi}\, F^{ab}\, F_{ab}\,,\label{emddilaton}
\eea
%%%%%%
where
%%%%%
\bea
R_{ab}=E_a{}^\mu\, E_b{}^\nu\, R_{\mu\nu}\,,\qquad
F_{ab}= E_a{}^\mu\, E_b{}^\nu\, F_{\mu\nu}\,,\qquad
\phi_a\equiv E_a{}^\mu\,\del_\mu\phi
\eea
%%%%%
and where 
$E^{a\mu}=\eta^{ab}\, E_b{}^\mu$, and $\varepsilon^{\mu\nu\rho\sigma}$
is the tensor density with components $\pm1$ and 0.

The perturbed
fields that are associated with the axial fluctuations and the polar
fluctuation are as follows:
%%%%%
\bea
\hbox{Axial}:&& (\delta\cB_0\,,\delta\cB_1\,,\delta \cB_2\,, \delta             F_{\u0\u3}\,,
\delta F_{\u1\u3}\,,\delta F_{\u2\u3})\,,\label{axiallist}\\
\hbox{Polar}:&& (\delta\nu\,,\delta\mu_1\,,\delta\mu_2\,,\delta\psi\,,
\delta\phi\,,\delta F_{\u0\u1}\,,\delta F_{\u0\u2}\,,\delta F_{\u1\u2})\,.
\label{polarlist}
\eea
%%%%%
As can be seen, the axial fields are of odd parity, undergoing a 
sign reversal under sending 
$\varphi\rightarrow -\varphi$, while the polar fields are of even parity 
and are unchanged under $\varphi\rightarrow -\varphi$. Since the background
Gibbons-Maeda black holes are themselves parity invariant, it follows
that the linearised equations 
describing the axial fluctuations will be completely decoupled from those 
describing the polar fluctuations.  In the next section, we study
the linearised fluctuations of the axial modes.  After that, we move on
to the more intricate analysis of the polar modes.

\section{The Axial Perturbations}

\subsection{The linearised field equations} 

    The axial perturbations involve the fluctuations listed
in (\ref{axiallist}), and these appear when substituting the expansions
(\ref{metricpert}) and (\ref{maxpert}) into the components 
$(\hbox{Ein})_{\u1\u3}$ and $(\hbox{Ein})_{\u2\u3}$ of the Einstein 
equations (\ref{tetradein}),
the component $(\hbox{Max})^\u3$ of the Maxwell equations (\ref{tetradmax}) 
and the components $(\hbox{Bianchi})^\u1$ and $(\hbox{Bianchi})^\u2$ of the
Bianchi identities (\ref{tetradbianchi}).

    It is useful to note that the relevant tetrad-frame components 
of the Ricci tensor $R_{ab}$, up to order $\epsilon$, are
%%%%%
\bea
R_{\u1\u3} &=& -\ft12 \epsilon\, e^{-2\psi-\nu-\mu_2}\, 
  \Big[\del_\theta(e^{3\psi+\nu-\mu_1-\mu_2}\, \cG_{12}) +
  \del_t(e^{3\psi-\nu-\mu_1+\mu_2}\, \cG_{01})\Big]\,,\nn\\
R_{\u2\u3} &=& \ft12 \epsilon\, e^{-2\psi-\nu-\mu_1}\, \Big[
  \del_r(e^{3\psi+\nu-\mu_1-\mu_2}\, \cG_{12}) -
\del_t(e^{3\psi-\nu+\mu_1-\mu_2}\, \cG_{02})\Big]\,,\label{Ric1323}
\eea
%%%%%
where $\cG=d\cB$ is the 2-form field strength of the Kaluza-Klein 
vector $\cB$ that appears in the general metric (\ref{genmet}).  (Note
that the components $\cG_{\alpha\beta}$ of the 2-form $\cG$ are the 
negatives of the quantities $Q_{\alpha\beta}$ defined in 
\cite{chanxant}.)
Note also that we are following the conventions stated earlier, of using
numerical indices $(\u0,\u1,\u2,\u3)$ to denote tetrad-index components,
and $(0,1,2,3)$ to denote coordinate-index components.  (In particular,
the Ricci tensor is given with tetrad components, whereas the
field strength $\cG$ has coordinate-index components.)  Explicit
partial derivatives $(\del_t,\del_r,\del_\theta)$ with respect to
the coordinates $(t,r,\theta)$ are interchangeable with 
$(\del_0,\del_1,\del_2)$.

  For now, since we are focusing only on the axial perturbations
we shall need just the $\u1$ and $\u2$ tetrad components 
of the Bianchi identities, which give
%%%%%
\bea
\del_\theta(e^{\psi+\nu}\, F_{\u0\u3}) -\del_t(e^{\psi+\mu_2}\, F_{\u2\u3}) 
&=& 0\,,\nn\\
\del_r(e^{\psi+\nu}\, F_{\u0\u3}) -\del_t(e^{\psi+\mu_1}\, F_{\u1\u3}) &=&0\,,
\label{bianchi12}
\eea
%%%%%
and the $\u3$ tetrad component of the Maxwell equations, which gives
%%%%%
\bea
&&\del_t(e^{\mu_1+\mu_2 +a\phi}\, F_{\u0\u3}) -
  \del_r(e^{\mu_2+\nu+a\phi}\, F_{\u1\u3}) -
  \del_\theta(e^{\mu_1+\nu+a\phi}\, F_{\u2\u3}) =\nn\\
&&- e^{\mu_2+a\phi+\psi}\, \cG_{01}\, F_{\u0\u1} -
 e^{\mu_1+a\phi+\psi}\, \cG_{02}\, F_{\u0\u2} +
  e^{\nu+a\phi+\psi}\, \cG_{12}\, F_{\u1\u2}\,.\label{max3}
\eea
%%%%%

  Note that the only non-vanishing tetrad components of the field strength
$F_{ab}$ in the background solution are given by $\bar F_{\u0\u1}=Q/r^2$.  
It follows that at the linearised level in fluctuations around the background,
all the metric functions in the Bianchi identities 
(\ref{bianchi12}) can be replaced by their background values.  Since these
are time-independent, the Bianchi identities (\ref{bianchi12}) imply
%%%%%
\bea
\del_t\, \delta F_{\u2\u3} = e^{-\bar\psi-\bar\mu_2}\, 
    \del_\theta(e^{\bar\psi+\bar\nu}\, \delta F_{\u0\u3})\,,\qquad
\del_t\, \delta F_{\u1\u3} = e^{-\bar\psi+\bar\nu}\,
    \del_r(e^{\bar\psi+\bar\nu}\, \delta F_{\u0\u3})\,.
\eea
%%%%%

   By the same token, in the $\u3$ component of the Maxwell equations 
given in (\ref{max3}), since the Kaluza-Klein field strength 
$\cG_{\alpha\beta}$ is also zero in the background, it follows that at
linear order we shall have
%%%%%
\bea
&&e^{-\bar\nu+\bar\mu_2 +a\bar\phi}\, \del_t\, \delta F_{\u0\u3} -
  \del_r(e^{\bar\mu_2+\bar\nu+a\bar\phi}\, \delta F_{\u1\u3}) -
  \del_\theta(e^{-\bar\nu+\bar\nu+a\bar\phi}\, \delta F_{\u2\u3}) =\nn\\
&&-e^{\bar\mu_2+a\bar\phi+\bar\psi}\, \delta \cG_{01}\, \bar F_{\u0\u1}\,.
\label{max3lin}
\eea
%%%%%

   Since all the metric functions in the background are independent of
$\theta$ except for $\bar\psi=\bar\mu_2 + \log\sin\theta$, we see that
the linearised Bianchi and Maxwell equations above reduce to
%%%%%
\bea
(\hbox{Bianchi})^\u1:&&  \del_t\,\delta F_{\u2\u3}= 
  e^{\bar\nu -\bar\mu_2}\, \fft1{\sin\theta}\, 
  \del_\theta(\sin\theta\, \delta F_{\u0\u3})\,,\nn\\
(\hbox{Bianchi})^\u2:&&  \del_t\,\delta F_{\u1\u3}=
  e^{\bar\nu-\bar\mu_2}\, \del_r(e^{\bar\nu+\bar\mu_2}\, 
\delta F_{\u0\u3})\,,\label{B1B2M3}\\
(\hbox{Max})^\u3:&& e^{-\bar\nu-\bar\mu_2}\, \del_t\, \delta F_{\u0\u3} -
e^{-2\bar\mu_2-a\bar\phi}\, 
 \del_r(e^{\bar\nu+\bar\mu_2 +a\bar\phi}\, \delta F_{\u1\u3}) -
e^{-2\bar\mu_2}\, \del_\theta \,\delta F_{\u2\u3}
= -\sin\theta\, \bar F_{\u0\u1}\, \delta\cG_{01}\,,\nn
\eea
%%%%%
where $\bar F_{\u0\u1}= \dfft{Q}{r^2}$ is the background value of the
field strength.

  Looking at the components $(\hbox{Ein})_{\u1\u3}$ and 
$(\hbox{Ein})_{\u2\u3}$ of the Einstein equation, it can be seen from
eqn (\ref{tetradein}) that to linear order in fluctuations we shall have
%%%%%%
\bea
(\hbox{Ein})_{\u1\u3}:&& R_{\u1\u3} +2e^{a\bar\phi}\, \bar F_{\u0\u1}\,
   \delta F_{\u0\u3}=0\,,\nn\\
(\hbox{Ein})_{\u2\u3}:&& R_{\u2\u3} =0\,.
\eea
%%%%%
(Note that there is no linear term from $\phi_\u1\, \phi_\u3$ in
$(\hbox{Ein})_{\u1\u3}$ (see eqn (\ref{tetradein})) even though
$\phi_\u1$ is non-zero in the background, because we are taking all
fluctuations to be independent of the azimuthal angle $\varphi$.) From
the expressions (\ref{Ric1323}) for the Ricci tensor components, we therefore
find at the linearised order
%%%%%
\bea
(\hbox{Ein})_{\u1\u3}:&& -e^{\bar\nu-\bar\mu_2}\, \fft1{\sin^2\theta}\,
\del_\theta(\sin^3\theta\, \delta \cG_{12}) - e^{-\bar\nu+\bar\mu_2}\,
\sin\theta\, \del_t\,\delta \cG_{01} +4 e^{a\bar\phi}\, \bar F_{\u0\u1}\,
\delta F_{\u0\u3}=0\,,\nn\\
(\hbox{Ein})_{\u2\u3}:&& e^{-2\bar\mu_2}\,
\del_r(e^{2\bar\nu+2\bar\mu_2}\, \delta\cG_{12}) -e^{-2\bar\nu}\, 
 \del_t\,\delta \cG_{02}=0\,.\label{Ein1323}
\eea
%%%%%%

   After some manipulations
of the perturbed field equations describing the axial fluctuations, 
it turns out that the
system can be reduced to just two second-order equations of motion for
the two perturbed fields
$\delta F_{\u0\u3}$ and $\delta \cG_{12}$.  To see this we follow 
a strategy analogous 
to that described in\cite{chanxant,chandra} for Reissner-Nordstr\"om.  
Thus we obtain the two coupled equations as follows:

\begin{itemize}

\item[(1)] Take the $\del_t$ derivative of $(\hbox{Max})^\u3$
in eqns (\ref{B1B2M3}), and then use $(\hbox{Bianchi})^\u1$ and 
$(\hbox{Bianchi})^\u2$ to eliminate the fluctuations $\delta F_{\u1\u3}$ and
$\delta F_{\u2\u3}$.  Then use $(\hbox{Ein})_{\u1\u3}$ in eqns 
(\ref{Ein1323}) to eliminate the fluctuation $\delta \cG_{01}$.  This
produces an equation involving only $\delta F_{\u0\u3}$ and $\delta\cG_{12}$,
with a wave operator $(-\del_t^2 + \del_{r_*}^2)$ 
acting on the fluctuation $\delta F_{\u0\u3}$.

\item[(2)] Write the $(\hbox{Ein})_{\u1\u3}$ equation as 
$\del_t\,\delta\cG_{01}=\cdots$ and the 
$(\hbox{Ein})_{\u2\u3}$ equation as $\del_t\,\delta\cG_{02}=\cdots$,
take the $\del_r$ derivative of the latter and subtract the $\del_\theta$
derivative of the former, and substitute these into the
right-hand side of the time derivative of $\del_t\,\delta\cG_{12}= 
\del_r\, \delta\cG_{02}-\del_\theta\,\delta\cG_{01}$ (which follows
from the Bianchi identity $d\cG=0$ for the field strength $\cG=d\cB$).  
The upshot is that we are left with an equation
involving only the fluctuations $\delta\cG_{12}$ and $\delta F_{\u0\u3}$,
with a wave operator $(-\del_t^2 + \del_{r_*}^2)$
acting on $\delta\cG_{12}$.

\end{itemize}

Implementing these procedures, and introducing new variables in place of 
$F_{\u0\u3}$ and $\cG_{12}$ by defining\footnote{The notation here
is that the minus superscript denotes axial perturbations, the
$1$ and $2$ subscripts denote that $H_1^-$ is associated with the fluctuation
of the spin-1 Maxwell field $F_{\u0\u3}$ and that $H_2^-$ is
associated with the fluctuation $\cG_{12}$ in the spin-2 metric field.}
%%%%%
\bea
H_1^- = -r e^{\bar\nu}\,\sin\theta\, \delta\,F_{\u0\u3}\,,\qquad
H_2^-= e^{2\bar\nu+\bar\mu_2}\, \sin^3\theta\, \delta\,\cG_{12}\,,
\eea
%%%%%
we find that the first itemised procedure above gives the equation
%%%%%
\bea
&&-\del_t^2\, H_1^- + \del_{r_*}^2\, H_1^-  + 
  \fft1{r^2}\, e^{2\bar\nu+a\bar\phi}\,\sin\theta\, 
 \del_\theta\Big(\fft1{\sin\theta}\, \del_\theta\, H_1^-\Big) -
\fft{4Q^2}{r^4}\, e^{2\bar\nu+a\bar\phi}\, H_1^- \nn\\
&&
  -e^{4\bar\nu}\,(a\bar\nu'\,\bar\phi' +\ft14 a^2\, {\phi'}^2 
         +\ft12 a \bar\phi'')\, H_1^-
 -\fft{Q}{r^3}\, e^{2\bar\nu+a\bar\phi}\, \fft1{\sin\theta}\,\del_\theta\,
         H_2^- =0\,.\label{axial1}
\eea
%%%%%
The second itemised procedure leads to the equation
%%%%%
\bea
&&-\del_t^2 \,H_2^- +  \del_{r_*}^2\, H_2^- + 
  \fft1{r^2}\, e^{2\bar\nu+a\bar\phi}\, \sin^3\theta\, \del_\theta\Big(
  \fft1{\sin^3\theta}\, \del_\theta\, H_2^-\Big) \nn\\
&&
 -e^{4\bar\nu}\, \Big(a \,\bnu'\,\bphi{'} + \ft14 a^2\, {\bphi{'}}^2
       + \ft12 a\,\bphi'' + \fft{2}{r^2}-\fft{2}{r}\, \bnu'
   -\fft{a}{r}\, \bphi' \Big)\, H_2^-\nn\\
&&
+\fft{4Q}{r^3} \, e^{2\bar\nu+a\bar\phi}\, \sin^3\theta\, \del_\theta\Big(
   \fft1{\sin^2\theta}\, H_1^-\Big) =0\,.\label{axial2}
\eea
%%%%%
Note that we have defined the $r_*$ coordinate such that
%%%%%
\bea
e^{2\bar\nu}\, \del_r = \del_{r_*}\,.
\eea
%%%%%
We have also made use of the background relations given in eqns 
(\ref{backgroundrels}) to write $\bar\mu_2$ in terms of $r$ and $\bar\phi$
using
%%%%
\bea
e^{\bar\mu_2}= r\, e^{-\ft12 a \bar\phi}\,.\label{backrel}
\eea
%%%%%
Additionally, we have written the background value of the Maxwell 
field strength as 
%%%%%
\bea
\bar F_{\u0\u1}= \fft{Q}{r^2}\,.
\eea
%%%%%

\subsection{Separating the axial equations}

  The next step is to separate the two equations (\ref{axial1}) and
(\ref{axial2}) into products of radial functions times $\theta$-dependent
angular functions times functions of time.  The time dependence is easily
factored out in the standard way, by taking all the fluctuations to have
time dependence $e^{-\im\omega t}$.  For the factoring out of the
angular dependence, we may follow a similar strategy to the one that 
was employed
in \cite{chanxant,xanth2}. Thus
we write
%%%%%
\bea
H_1^-(t,r,\theta) = 3 e^{-\im\omega t}\, \wtd H_1^-(r)\, 
C_{\ell+1}^{-\ft12}(\cos\theta)\,,\qquad 
H_2^-(t,r,\theta) = e^{-\im\omega t}\, H_2^-(r)\, 
C_{\ell+2}^{-\ft32}(\cos\theta)\,,\label{gegenbauersubs}
\eea
%%%%%
where $C_n^{\alpha}(x)$ are the Gegenbauer 
polynomials.
Using standard properties of these polynomials,
one can see that the angular dependence
factors out and one is left with purely radial equations.\footnote{The
Gegenbauer polynomials $C_n^{\alpha}(x)$ can be defined by the
generating function $(1-2x\, t +t^2)^{-\alpha} =
\sum_{n\ge 0} t^n\, C_n^{\alpha}(x)$, and they satisfy the 
Gegenbauer equation $(1-x^2)\, C_n^\alpha{}'' -(2\alpha+1)\,x\,
C_n^\alpha{}' + n\,(n+2\alpha)\, C_n^\alpha=0$.  It is useful
also to note that $C_n^\alpha{}'= 2\alpha\, C_{n-1}^{\alpha+1}$.
They are related to certain
specialisations of the spin-weighted spherical harmonics 
$_s Y_{\ell, m}(\theta,\varphi)$, with $C_n^{\alpha}(\cos\theta)
= (\hbox{const})\,  _s Y_{\ell, 0}(\theta,0)\,(\sin\theta)^s$ for
$\alpha=\ft12 - s$ and $n=\ell+s$.  Thus $C_{\ell+1}^{-\ft12}(\cos\theta)$
is proportional to $_{1}Y_{\ell,0}(\theta,0)\, \sin\theta$ and 
$C_{\ell+2}^{-\ft32}(\cos\theta)$ is proportional to 
   $_{2}Y_{\ell,0}(\theta,0)\, \sin^2\theta$. }

  The two radial equations that result from plugging the substitutions 
(\ref{gegenbauersubs}) into eqns (\ref{axial1}) and (\ref{axial2}) are
%%%%%
\bea
(\omega^2+\del_{r_*}^2)\, \wtd H_1^-(r) + T_1\, \wtd H_1^-(r) 
   + S_1\, H_2^-(r) &=&0
\,,\nn\\
(\omega^2+\del_{r_*}^2)\, H_2^-(r) + T_2\, H_2^-(r) + S_2\, \wtd H_1^-(r) &=&0
\,,\label{Hm1Hm2eqns}
\eea
%%%%%
where
%%%%%
\bea
T_1 &=& -  e^{4\bnu} \, \Big(a \,\bnu'\,\bphi{'} + \ft14 a^2\, {\bphi{'}}^2 
+ \ft12 a\,\bphi''\Big)
  -\fft1{r^2}\, e^{2\bnu+a\bphi}\,
\Big(\fft{4 Q^2}{r^2}  + \ell(\ell+1)\Big)\,,\nn\\
S_1 &=& -\fft{Q}{r^3}\, e^{2\bnu+a\bphi}\,,\nn\\
T_2 &=& -  e^{4\bnu} \,\Big(a \,\bnu'\,\bphi{'} + \ft14 a^2\, {\bphi{'}}^2
       + \ft12 a\,\bphi'' + \fft{2}{r^2}-\fft{2}{r}\, \bnu' 
   -\fft{a}{r}\, \bphi' \Big) -\fft1{r^2}\, e^{2\bnu+a\bphi}\,
  \Big(\ell(\ell+1) -2\Big)\,,\nn\\
S_2 &=& -\fft{4 Q}{r^3}\, e^{2\bnu+a\bphi} \,\Big(\ell(\ell+1)-2\Big)\,.
\eea
%%%%% 

  Noting that the first three terms in the bracket multiplying $e^{4\bnu}$
in the expression for $T_2$ are the same as the three terms  in the
bracket multiplying $e^{4\bnu}$ in the expression for $T_1$, we take the
extra three terms in the $T_2$ expression and write them as
%%%%%
\bea
e^{4\bnu}\, \Big(\fft{2}{r^2}-\fft{2}{r}\, \bnu'   
   -\fft{a}{r}\, \bphi' \Big) &=& e^{2\bnu+a\bphi}\, e^{2\bnu-a\bphi}\,
    \Big(\fft{2}{r^2}-\fft{2}{r}\, \bnu'
   -\fft{a}{r}\, \bphi' \Big) \nn\\
&=& e^{2\bnu+a\bphi}\, \Big(1-\fft{\rps}{r}\Big)\Big(1-\fft{\rms}{r}\Big)
  \Big(\fft{2}{r^2}-\fft{2}{r}\, \bnu'
   -\fft{a}{r}\, \bphi' \Big)\,.
\eea
%%%%%
(See (\ref{bhsol}) and (\ref{backgroundrels}).)  Making further use
of the expressions for the metric functions and $\bphi$, we find that the
equations (\ref{Hm1Hm2eqns}) can be written as
%%%%%
\bea
(\omega^2+\del_{r_*}^2 + T)\,  H_1^- +\fft{1}{r^3}\, e^{2\bnu+a\bphi}\,
\Big( - C\,  H_1^- -2\mu\, Q\, H_2^-\Big)&=&0\,,\nn\\
(\omega^2+\del_{r_*}^2 + T)\, H_2^- +\fft{1}{r^3}\, e^{2\bnu+a\bphi}\,
\Big(  C\, H_2^- -2\mu\, Q\, H_1^-\Big)&=&0\,,\label{H1H2res}
\eea
%%%%%
where we have defined
%%%%%%
\bea
T &=& -a\, e^{4\bnu}\, \Big(\bnu'\,\bphi{'} + \ft14 a\, {\bphi{'}}^2
       + \ft12 \bphi''\Big) +\fft1{r^2}\, e^{2\bnu+a\bphi}\,
\Big[\fft{C}{r} -\ell(\ell+1) - \fft{4 Q^2}{r^2}\Big]\,,\\
H_1^- &=& 2\mu\, \wtd H_1^-\,, 
\eea
%%%%%
and the constants
%%%%%
\bea
C= \fft32 (\rps + \rms)-\fft{2 a^2\,\rms}{1+a^2}\,,\qquad
\mu = \sqrt{\ell(\ell+1)-2}=\sqrt{2n} \,.\label{Cmudef}
\eea
%%%%%
Recall that we are assuming $\ell\ge2$, and so $\mu$ is real and positive. 
(Note that some rather non-trivial ``conspiracies'' occur in order to
arrive at the relatively simple expressions in eqns (\ref{H1H2res}).)

   Finally, we see that by defining the linear combination
%%%%%
\bea
Z^- = \alpha\, H_1^-  + \beta\, H_2^-\,,
\eea
%%%%%
then for 
%%%%%
\bea
\fft{\beta}{\alpha} = \fft{ -C \pm\sqrt{C^2+ 4\mu^2 Q^2}}{2\mu\, Q}=
\fft{2\mu\, Q}{C\pm\sqrt{C^2+ 4\mu^2 Q^2}}
\label{albesol}
\eea
%%%%%%
we get decoupled equations for functions
$Z^-_1$ and $Z^-_2$ (corresponding to the plus sign and minus sign choices
in eqn (\ref{albesol})), obeying
%%%%%
\bea
(\del_{r_*}^2 +\omega^2)\, Z^-_i = V_i^-\, Z^-_i\,,\qquad i=1, 2
\eea
%%%%%
with 
%%%%%
\bea
V^-_{1,2}= -T \pm  e^{2\bnu+a\bphi}\, \fft{
\sqrt{C^2+ 4\mu^2\, Q^2}}{r^3} = -T \pm e^{2\bnu-2\bmu_2}\,\fft{ 
       \sqrt{C^2+ 4\mu^2\, Q^2}}{r} \,.\label{VM12}
\eea
%%%%%

\section{The Polar Perturbations}

\subsection{Linearised polar fluctuations}

  At the linearised level, these involve the fluctuations listed in eqn 
(\ref{polarlist}).  The relevant field equations that govern these fluctuations
are the linearised fluctuations of the $(\hbox{Ein})_{\u0\u1}$
$(\hbox{Ein})_{\u0\u2}$, $(\hbox{Ein})_{\u1\u1}$,
$(\hbox{Ein})_{\u1\u2}$ and $(\hbox{Ein})_{\u3\u3}$ Einstein equations 
in eqn (\ref{tetradein}); the Maxwell equations $(\hbox{Max})^{\u1}$ and 
$(\hbox{Max})^{\u2}$ in eqn (\ref{tetradmax}); the Bianchi 
identity $(\hbox{Bianchi})^\u3$ in eqn (\ref{tetradbianchi}); and
the dilaton equation (\ref{emddilaton}).

  From the expansions (\ref{metricpert}), we find that up to linear order in
$\ep$,
%%%%%
\bea
R_{\u0\u1}&=& \ep\, \delta R_{\u0\u1}\,,\nn\\
R_{\u0\u2}&=& \ep\, \delta R_{\u0\u2}\,,\nn\\
R_{\u1\u2}&=& \ep\, \delta R_{\u1\u2}\,,\nn\\
R_{\u1\u1}&=& -e^{2\bar\nu}\, \Big[2{\bmu_2{}'}^2 + 2\bar\mu_2'\,\bar\nu' +
2{\bar\nu{}'}^2 +\bar\nu''\Big] + \ep\, \delta R_{\u1\u1}\,,\nn\\
R_{\u3\u3}&=& e^{-2\bmu_2} - e^{2\bnu}\,
  (2 {\bmu_2{}'}^2 + 2 \bmu_2'\, \bnu' + \bmu_2'') +
   \ep\,\delta R_{\u3\u3}\,,\label{Rexpansions}
\eea
%%%%%
where
%%%%%
\bea
\delta R_{\u0\u1}&=& - \del_t\,\Big[\del_r\, (\delta\mu_2+\delta\psi) + 
 (\bar\mu_2'-\bar\nu')(\delta\mu_2 +\delta\psi) - 
 2\bar\mu_2'\,\delta\mu_1\Big]\,,\nn\\
\delta R_{\u0\u2}&=& e^{-\bar\nu-\bar\mu_2}\, \del_t\,
\Big[ (\delta\mu_2-\delta\psi)\,\cot\theta-
    \del_\theta\, (\delta\mu_1+\delta\psi)
\Big]\,,\nn\\
\delta R_{\u1\u2}&=& e^{\bar\nu-\bar\mu_2}\, \Big[-\del_r\del_\theta\,
 (\delta\nu+\delta\psi) + \del_r\, (\delta\mu_2-\delta\psi)\, \cot\theta +
  \bar\mu_2'\, \del_\theta\, (\delta\mu_1 +\delta\nu) \nn\\
&&\qquad\qquad +
\bar\nu'\, \del_\theta\, (\delta\mu_1-\delta\nu)\Big]\,,\nn\\
\delta R_{\u1\u1}&=& \Big\{e^{2\bar\nu}\, 
         \Big[ (4{\bar\mu{}_2'}^2 + 4 \bar\mu_2'\,\bar\nu' +
     4 {\bar\nu{}'}^2 + 4 \bar\mu_2'' + 2 \bar\nu'')\, \delta\mu_1 + 
  (2\bar\mu_2'+\bar\nu')\, \del_r\, (\delta\mu_1-\delta\mu_2 -\delta\psi) \nn\\
&&\qquad\qquad -
   3 \bar\nu'\, \del_r\, \delta\nu - \del_r^2\, 
(\delta\mu_2+\delta\nu+\delta\psi)\Big] -
  e^{-2\bar\mu_2}\,\fft1{\sin\theta}\, \del_\theta\,
 (\sin\theta\, \del_\theta\,\delta\mu_1)\nn\\
&&\qquad + 
 e^{-2\bar\nu}\, \del_t^2\, \delta\mu_1 \Big\}\,,\nn\\
\delta R_{\u3\u3}&=&
\Big\{ e^{2\bnu}\, \Big[ -\del_r^2\, \delta\psi + (4 {\bmu_2{}'}^2 
   + 4 \bmu_2'\, \bnu' + 2 \bmu_2'')\, \delta\mu_1 +
   \bmu_2'\, \del_r\,(\delta\mu_1 - \delta\mu_2 -\delta\nu)\nn\\
&&\qquad\qquad -
  (3\bmu_2' + 2 \bnu')\, \del_r\, \delta\psi \Big]\nn\\
&&
 \ \  -
  e^{-2\bmu_2}\, \Big[2\delta\mu_2 + \del_\theta^2\, \delta\psi +
\cot\theta\, \del_\theta\, 
  (\delta\mu_1 -\delta\mu_2 + \delta\nu + 2 \delta\psi)\Big] +
e^{-2\bnu}\, \del_t^2\, \delta\psi\Big\}\,.\label{deltaR}
\eea
%%%%%
Note that the diagonal components $R_{\u1\u1}$ and $R_{\u3\u3}$ listed
here have terms at order $\ep^0$ as well as at order $\ep$ (see eqns 
(\ref{Rexpansions})).  The $\ep^0$ 
terms form part of the background black hole solution itself.

  As can be verified by expanding the Einstein equations (\ref{tetradein})
up to order $\ep$, the equations $(\hbox{Ein})_{\u0\u1}$
$(\hbox{Ein})_{\u0\u2}$, $(\hbox{Ein})_{\u1\u1}$,
$(\hbox{Ein})_{\u1\u2}$ and $(\hbox{Ein})_{\u3\u3}$ that are relevant for 
the polar perturbations, give
%%%%%
\bea
\delta R_{\u0\u1} -\ft12 \bar\phi'\, \del_t\, \delta\phi&=&0\,,\nn\\
\delta R_{\u0\u2} + \fft{2 Q}{r^2}\, e^{a \bar\phi}\, 
       \delta F_{\u1\u2}&=&0\,,\nn\\
\delta R_{\u1\u2} -\ft12 e^{\bnu-\bmu_2}\, \bar\phi'\, 
                      \del_\theta\, \delta\phi + 
   \fft{2Q}{r^2}\, e^{a\bar\phi}\, \delta F_{\u0\u2} &=&0\,,\nn\\
\delta R_{\u1\u1} + e^{2\bnu}\, ({\bar\phi{}'}^2\, \delta\, \mu_1 -
\bar\phi'\, \del_r\, \delta\phi) +
 e^{a\bar\phi}\, \Big(\fft{a\, Q^2}{r^4}\,\delta\phi + \fft{2Q}{r^2}\,
\delta F_{\u0\u1}\Big) &=&0\,,\nn\\
\delta R_{\u3\u3} - e^{a\bar\phi}\,\Big(\fft{a\,Q^2}{r^4}\,  \delta\phi
- \fft{2 Q}{r^2}\, \delta F_{\u0\u1}\Big) &=&0\,,\label{einstpolar}
\eea
%%%%%
where the varied Ricci tensors are given in eqns (\ref{deltaR}).

  The remaining equations that we need in order to determine the polar
perturbations are the $\u1$ and $\u2$ components of the Maxwell 
equations (\ref{tetradmax}),
the $\u3$ component of the Bianchi identities (\ref{tetradbianchi}) and
the dilaton equation of motion (\ref{emddilaton}).  Working to
linear order in $\ep$, these give
%%%%%
\bea
(\hbox{Max})^{\u1}:&& e^{\bnu}\, \fft1{\sin\theta}\, \del_\theta(
\sin\theta\, \delta F_{\u1\u2})+ e^{\bmu_2}\, \Big[\fft{Q}{r^2}\, 
\del_t(\delta\mu_2 + a\, \delta\phi +\delta\psi) + 
 \del_t\,\delta F_{\u0\u1}\Big] =0\,,\nn\\
(\hbox{Max})^{\u2}:&& \del_r\,( e^{\bnu+\bmu_2+a\, \bar\phi}\, 
 \delta F_{\u1\u2}) -
e^{-\bnu+\bmu_2+a\, \bar\phi}\, \del_t\, \delta F_{\u0\u2} =0\,,\nn\\
(\hbox{Bianchi})^{\u3}:&&
 \del_\theta\Big[\delta F_{\u0\u1} + \fft{Q}{r^2}\, 
  (\delta\mu_1+\delta\nu)\Big] -\del_r (e^{\bmu_2+\bnu}\, \delta F_{\u0\u2})
 + e^{\bmu_2-\bnu}\, \del_t\, \delta F_{\u1\u2}=0\,,\label{Max12B3}
\eea
%%%%%
and, from the dilaton equation,
%%%%%
\bea
&&e^{2\bnu}\, \Big[\del_r^2\, \delta\phi + 
 2 (\bmu_2'+\bnu')\, \del_r\, \delta\phi +
\bar\phi'\, \del_r(\delta\psi-\delta\mu_1+\delta\mu_2+\delta\nu) -
(2 \bar\phi'' + 4 \bmu_2'\, \bar\phi' + 4 \bnu'\, \bar\phi')\,
\delta\mu_1\Big]\nn\\
&& + e^{-2\bmu_2}\, \fft{1}{\sin\theta}\del_\theta(
\sin\theta\,\del_\theta\,\delta\phi) - 
  e^{-2\bnu}\, \del_t^2\, \delta\phi + 
 e^{a\bar\phi}\, \Big[ \fft{2 a^2\,Q^2}{r^4}\, \delta\phi +
   \fft{4 a Q}{r^2}\, \delta F_{\u0\u1}\Big] =0\,.\label{dileom0}
\eea
%%%%%

\subsection{Separation of variables}

  We proceed to separate variables by making substitutions similar
to those employed for Reissner-Nordstr\"om in \cite{chanxant,xanth2}.  
Thus, with the understanding that $P_\ell$ means $P_\ell(\cos\theta)$, 
we write
%%%%%
\bea
\delta\nu &=& e^{-\im\omega t}\, 
 N(r)\, P_\ell\,,\qquad 
\delta\mu_1=e^{-\im\omega t}\, L(r)\, P_\ell\,,\nn\\
\delta\mu_2&=& e^{-\im\omega t}\,\big[ T(r)\, P_\ell + 
                                V(r)\, \del_\theta^2 P_\ell\big]\,,\nn\\
\delta\psi &=& e^{-\im\omega t}\,\big[ T(r)\, P_\ell + 
                                V(r)\, \cot\theta\, 
                               \del_\theta P_\ell\big]\,,\nn\\
\delta F_{\u0\u1} &=& -\fft{r^2\, e^{2\bnu-a\bar\phi}}{2Q}\, e^{-\im\omega t}\,
B_{\u0\u1}\, P_\ell\,,\nn\\
\delta F_{\u0\u2} &=& \fft{r\, e^{\bnu-\ft12 a\bar\phi}}{2Q}\, 
e^{-\im\omega t}\,
B_{\u0\u2}\, \del_\theta P_\ell\,,\nn\\
\delta F_{\u1\u2} &=& -\im\omega\, \fft{r\, e^{-\bnu-\ft12a\bar\phi}}{2Q}\, 
e^{-\im\omega t}\,
B_{\u1\u2}\, \del_\theta P_\ell\,,\nn\\
\delta\phi &=& e^{-\im\omega t}\,\Phi(r)\, P_\ell\,.\label{redefs}
\eea
%%%%%
(We chose the powers of $e^{a\bar\phi}$ appearing in the expressions 
for the field strength fluctuations in order simplify the subsequent
equations as much as possible.)

    Note that since the Legendre polynomial $P_\ell$ satisfies
%%%%%
\bea
\fft{1}{\sin\theta}\, \del_\theta\big(\sin\theta\, \del_\theta P_\ell\big)
  + \ell(\ell+1)\, P_\ell=0\,,
\eea
%%%%%
it follows that
%%%%%
\bea
\delta\mu_2+\delta\psi= e^{-\im\omega t}\,\big[ 2 T(r) -\ell(\ell+1)\, V(r)
                                          \big]\, P_\ell\,.
\eea
%%%%%   

 Suppressing the factors of $e^{-\im\omega t}$ after taking $t$ derivatives, 
since $e^{-\im\omega t}$ will be common to all terms in all the equations,
we then have from the equations (\ref{deltaR}) that
%%%%%
\bea
\delta R_{\u0\u1} &=& \im\omega\, \Big[2T' -\ell(\ell+1)\, V' +
   (\bmu_2'-\bnu')[2 T -\ell(\ell+1) V] -2\bmu_2'\, L\Big]\, P_\ell\,,\nn\\
\delta R_{\u0\u2} &=& \im\omega\, e^{-\bnu-\bmu_2}\, \big[L+T-V\big]\, 
  \del_\theta P_\ell\,,\nn\\
\delta R_{\u1\u2} &=& e^{\bnu-\bmu_2}\, \Big[
 (\bmu_2'+\bnu') L + (\bmu_2' -\nu') N -N'-T'+V'\Big]\, \del_\theta P_\ell\,,
\nn\\
\delta R_{\u1\u1} &=& \Big\{ e^{2\bar\nu}\,
         \Big[ (4{\bar\mu{}_2'}^2 + 4 \bar\mu_2'\,\bar\nu' +
     4 {\bar\nu{}'}^2 + 4 \bar\mu_2'' + 2 \bar\nu'')\, L +
  (2\bar\mu_2'+\bar\nu')\, (L'-2T' + \ell(\ell+1)\, V') \nn\\
&&\qquad\qquad -
   3 \bar\nu'\, N' -
(N'' + 2 T'' - \ell(\ell+1)\, V'')\Big] +
  e^{-2\bar\mu_2}\,\ell(\ell+1)\, L \nn\\
&&\qquad 
-\omega^2\,  e^{-2\bar\nu}\, L \Big\}\, P_\ell\,,\nn\\
\delta R_{\u3\u3} &=& \Big\{ e^{2\bnu}\, \Big[ 
 (4{\bmu_2{}'}^2 + 4 \bmu_2'\, \bnu' +2\bmu_2'')L +
   \bmu_2'\, (L'-N') -(4\bmu_2'+2\bnu')\, T' +\ell(\ell+1)\, \bmu_2'\, V'
  -T''\Big]\nn\\
&&
 +e^{-2\bmu_2}\, (\ell+2)(\ell-1)\, T - 
\omega^2 \,e^{-2\bnu}\, T\Big\}\, P_\ell \nn\\
&&- \Big\{e^{2\bnu}\, [V'' +2(\bmu_2'+\bnu')\, V']  +
   e^{-2\bmu_2}\, (L+N) + \omega^2 \, e^{-2\bnu}\, V\Big\}\, \cot\theta\,
  \del_\theta P_\ell\,.
\eea
%%%%%
 
The equations for the fluctuations become
%%%%%
\bea
(\hbox{Ein})_{\u0\u1}:&& \Big(\del_r + \bmu_2'-\bnu'\Big)
 \big[2T -\ell(\ell+1) \, V\big] - 2\bmu_2'\, L +
              \ft12 \bar\phi'\, \Phi=0\,,\label{rein01}\\
(\hbox{Ein})_{\u0\u2}:&& L+T-V -
      B_{\u1\u2}=0\,,\label{rein02}\\
(\hbox{Ein})_{\u1\u2}:&&  T'-V'+N' - (\bmu_2'+\bnu')\, L -
   (\bmu_2'-\bnu')\, N +\ft12 \bar\phi'\, \Phi -
      B_{\u0\u2}=0\,,\label{rein12}\\
(\hbox{Ein})_{\u1\u1}:&&
2\bmu_2'\, N' + (\bmu_2'+\bnu') \big[2T'-\ell(\ell+1)\, V'\big]
  +(-2 {\bmu_2{}'}^2 -4 \bmu_2'\,\bnu' +\ft12 {\bar\phi{}'}^2)\, L\nn\\
&&
- e^{-2\bmu_2-2\bnu}\, \Big[ \ell(\ell+1)\, N + (\ell^2 +\ell-2)\, T)\Big]
+\omega^2\, e^{-4\bnu}\, \big[2T-\ell(\ell+1)\, V\big]
\nn\\
&&-  B_{\u0\u1} -\ft12 \bar\phi'\,\Phi'
+ \fft{a Q^2}{r^4}\,e^{-2\bnu+a\bar\phi} \,\Phi=0  \,, \label{rein11}\\
&& \nn\\
(\hbox{Ein})_{\u3\u3}:&&N''+T''-\ell(\ell+1)\, V'' +
  (\bmu_2'+ 3\bnu')\, N' +(\bmu_2'+\bnu')\, \big[2 T' - 2\ell(\ell+1)\, V'
    -L'\big] \nn\\
&&-2(\mu_2'' + \bnu'' + \bmu_2'{}^2+2 \bnu'{}^2 +
     2\bmu_2'\,\bnu'+\ft14 {\bar\phi{}'}^2)\, L 
  -\ell(\ell+1)\, e^{-2\bmu_2-2\bnu}\,(L+N)\nn\\
&& +
\omega^2\, e^{-4\bnu}\, \big[L+T-\ell(\ell+1)\, V\big]+  B_{\u0\u1} +
\ft12\bar\phi'\, \Phi' - \fft{a Q^2}{r^4}\, e^{-2\bnu+a\bar\phi}\, \Phi=0\,,
\label{rein331}\\
&&\nn\\
&& V'' +2(\bmu_2'+\bnu')\, V'  +
   e^{-2\bmu_2-2\bnu}\, (L+N) + \omega^2 \, e^{-4\bnu}\, V=0\,,\label{rein332}
\eea
%%%%%
and the Maxwell equations and Bianchi identity in eqns 
(\ref{Max12B3}) give
%%%%%
\bea
(\hbox{Max})^{\u1}:&&
\ell(\ell+1)\,  B_{\u1\u2} +  e^{2\bnu+2\bmu_2}\,  B_{\u0\u1} \nn\\
&&- 2 e^{2\bmu_2+ a\bar\phi}\, \fft{Q^2}{r^4}\,
\big[2T -\ell(\ell+1) V + a  \Phi\big] =0\,,\label{rmax1}\\
(\hbox{Max})^{\u2}:&& 
\fft1{r^2}\, \del_r\big( r^2\,  B_{\u1\u2}\big) -
  B_{\u0\u2}=0\,,\label{rmax2}\\
(\hbox{Bianchi})^{\u3}:&& 
e^{2\bnu-a\bar\phi}\, B_{\u0\u1} \! + \fft{1}{r^2}\, 
 \del_r\big(r^2\, e^{2\bnu-a\bar\phi}\,  B_{\u0\u2}\big)
 +
\omega^2\, e^{-2\bnu-a \bar\phi}\, B_{\u1\u2} 
\nn\\
&& -\fft{2Q^2}{r^4}\,(N+L) =0\,.\label{rbianchi3}
\eea
%%%%%
Finally, the dilaton equation (\ref{dileom0}) gives
%%%%%
\bea
&&e^{2\bnu}\, \Big[\Phi'' +
 2 (\bmu_2'+\bnu')\, \Phi' +
\bar\phi'\, \big[2T'- \ell(\ell+1)\,V' +N'-L'\big]
\nn\\
&&\qquad -(2 \bar\phi'' + 4 \bmu_2'\, \bar\phi' + 4 \bnu'\, \bar\phi')\, L
-2a\,  B_{\u0\u1}\Big]\nn\\
&& -\ell(\ell+1)\,  e^{-2\bmu_2}\,\Phi  +
  \omega^2\, e^{-2\bnu}\,\Phi  +
 \fft{2 a^2\,Q^2}{r^4}\, e^{a\bar\phi}\, \Phi =0\,.\label{rdileom}
\eea
%%%%%

  It is worth remarking that eqns (\ref{rein331}) and (\ref{rein332}),
which arise respectively from the terms proportional to $P_\ell(\cos\theta)$
and $\del_\theta P_\ell(\cos\theta)$ in the perturbed $(\hbox{Ein})_{\u3\u3}=0$
equation, are in fact derivable from the other equations contained in
eqns (\ref{rein01})--(\ref{rdileom}), and therefore we shall not need to make 
use of them directly in what follows.  Note also that we used the relation 
(\ref{backrel}), which is obeyed by the background fields, in various
places when we obtained eqns (\ref{rein01})--(\ref{rdileom}).

\subsection{First-order equations}

   In the Reissner-Nordstr\"om case, the complete system of polar 
radial equations 
following from the Einstein-Maxwell equations could be reduced down to
five first-order equations for five independent variables 
\cite{xanth2,chandra}.
It turns out that in our EMD case we can reduce the
complete system of polar radial equations to five first-order equations and
a second-order equation (the dilaton equation).  For convenience, we can
then 
rewrite the dilaton equation in terms of two first-order equations, so that
we end up with seven first-order equations in seven independent variables.

  To do this, we first note that eqn (\ref{rein02}) can be used to solve 
algebraically for $T$, and that eqn (\ref{rmax1}) can be used to solve
algebraically for $B_{\u0\u1}$.  Defining also the auxiliary variable
$\Psi=\Phi'$, we are then left with the following seven independent
variables:
%%%%%
\bea
\{ N,L, X, B_{\u1\u2}, B_{\u0\u2}, \Phi,\Psi\}\,,\label{sevenvar}
\eea
%%%%%
where, as in \cite{chanxant,xanth2}, we have defined
%%%%%
\bea
X= n\, V\,,\qquad n=\ft12\ell(\ell+1)-1\,.\label{Xndef}
\eea
%%%%%
The equations (\ref{rein01}), (\ref{rein12}), (\ref{rein11}), (\ref{rmax2}),
(\ref{rbianchi3}) then yield first-order equations for 
$\{N,L, X, B_{\u1\u2}, B_{\u0\u2}\}$, and the dilaton equation
(\ref{rdileom}), together with the definition $\Psi=\Phi'$, yields 
first-order equations for $\Phi$ and $\Psi$.  
This complete system of seven first-order equations is then given by
%%%%%
\bea
N' &=& \td a\, N + \td b\, L + 
\td c\, (X- B_{\u1\u2}) 
+ \fft1{4\bmu_2'}\, \Big[\bar\phi'\, \Psi + (\bmu_2' +\bnu')\, \bar\phi'\,\Phi
+ \fft{2 a Q^2}{r^2}\, e^{-2\bmu_2-2\bnu}\, \Phi\Big]\,,\nn\\
L' &=& (\td a-\bmu_2' +\bnu')\, N +
      (\td b-\bmu_2' -\bnu' )\, L +
   \td c\, (X- B_{\u1\u2}) - \fft{2}{r}\, B_{\u1\u2} 
  \nn\\
&& 
+\fft1{4\bmu_2'}\, \Big[\bar\phi'\, \Psi+(3\bmu_2' +\bnu')\, \bar\phi'\,\Phi
+ \fft{2 a Q^2}{r^2}\, e^{-2\bmu_2-2\bnu}\, \Phi\Big]\,,\nn\\
X'&=& -(\td a-\bmu_2'+\bnu')\, N -
  (\td b+\bmu_2' -2\bnu' )\, L -
(\td c+\bmu_2'-\bnu')\, (X-B_{\u1\u2}) + B_{\u0\u2} \nn\\
&&
-
  \fft1{4\bmu_2'}\, \Big[\bar\phi'\, \Psi+(2\bmu_2' +\bnu')\, \bar\phi'\,\Phi
+ \fft{2 a Q^2}{r^2}\, e^{-2\bmu_2-2\bnu}\, \Phi\Big]\,,\nn\\
B_{\u1\u2}' &=& B_{\u0\u2} - \fft{2}{r}\, B_{\u1\u2}\,,\nn\\
B_{\u0\u2}' &=& \fft{2 Q^2}{r^2}\, e^{-2\bmu_2-2\bnu}\, 
(N + 3 L + 2X -a\, \Phi) +
   \Big[ \fft{2(n+1) r^2-4Q^2}{r^2}\, e^{-2\bmu_2-2\bnu} -
   \omega^2\, e^{-4\bnu}\Big]\, B_{\u1\u2}\nn\\
&& 
  -\Big[ 2\bnu' -a\,\bar \phi' +\fft2{r}\Big]\, B_{\u0\u2}\,,\nn\\
\Phi' &=& \Psi\,,\nn\\
\Psi' &=& (\bnu'-\bmu_2')\, \bar\phi'\, N +
   \Big[(5\bnu'-\bmu_2')\,\bar\phi' + 2\bar\phi''-\fft{8a Q^2}{r^2}\,
   e^{-2\bmu_2-2\bnu}\Big]\, L\nn\\
&&
 + \Big[(2\bnu'-2\bmu_2')\,\bar\phi' -\fft{8a Q^2}{r^2}\,
   e^{-2\bmu_2-2\bnu}\Big]\,  X\nn\\
&& +\Big[\Big(2\bmu_2'-2\bnu'-\fft2{r}\Big)\,\bar\phi' +\fft{8a Q^2}{r^2}\,
   e^{-2\bmu_2-2\bnu} -4(n+1)a\, e^{-2\bmu_2-2\bnu}\Big]\, B_{\u1\u2} 
\nn\\
&&
+\Big[ {\bar\phi{}'}^2+ 2(n+1)\, e^{-2\bmu_2-2\bnu} -\omega^2\, e^{-4\bnu} 
+\fft{2a^2\,  Q^2}{r^2}\, e^{-2\bmu_2-2\bnu}\Big]\, \Phi -
2(\bmu_2'+\bnu')\, \Psi\,,\label{7firstorder}
\eea
%%%%%
where
%%%%%
\bea
\td a &=& \fft{n+1}{\bmu_2'}\, e^{-2\bmu_2-2\bnu}\,,\nn\\
\td b&=&  -\bmu_2' +\bnu' -\fft{n}{\bmu_2'}\, e^{-2\bmu_2-2\bnu} +
     \fft{{\bnu{}'}^2}{\bmu_2'} + \fft{\omega^2}{\bmu_2'}\, e^{-4\bnu} -
   \fft{2 Q^2}{\bmu_2'\, r^2}\, e^{-2\bmu_2-2\bnu}- 
  \fft{{\bar\phi{}'}^2}{4 \bmu_2'}\,,\nn\\
\td c &=& -\bmu_2'  +\fft{1}{\bmu_2'}\, e^{-2\bmu_2-2\bnu} +
     \fft{{\bnu{}'}^2}{\bmu_2'} + \fft{\omega^2}{\bmu_2'}\, e^{-4\bnu} -
   \fft{2 Q^2}{\bmu_2'\, r^2}\, e^{-2\bmu_2-2\bnu}\,.\label{abctilde}
\eea
%%%%%

\section{Diagonalisation Of The Polar Modes For EMD Black Holes}

  Guided by the way the polar modes in the Reissner-Nordstr\"om background
were found in \cite{xanth2,chandra}, we now make an ansatz for the
EMD case.  We begin by defining 
%%%%%
\bea
Z^+= f_1\, X + f_2\, (L+X-B_{\u1\u2}) + f_3\, B_{\u1\u2} + 
  f_4\, \Phi + f_5\, \Psi\,,\label{Zans}
\eea
%%%%%
where the $f_i$ are all functions of $r$, to be determined.  We then
require that $Z^+$ should satisfy
%%%%%
\bea
(\Lambda^2 + \omega^2)Z^+ - g\, Z^+=0\,,\label{Zeqn}
\eea
%%%%%
where 
%%%%%
\bea
\Lambda^2 = \del_{r_*}^2\,,\qquad \hbox{where}\quad
\del_{r_*}= e^{2\bar\nu}\, \del_r\,,
\eea
%%%%%
and $g$ is another function of $r$ to be determined.

   By substituting $Z^+$ defined in eqn (\ref{Zans}) into eqn (\ref{Zeqn}), 
and making repeated use of the set of the seven first-order equations 
(\ref{7firstorder}) for the derivatives 
$(N',L',X',B_{\u1\u2}', B_{\u0\u2}',\Phi',\Psi')$, 
eqn (\ref{Zeqn}) can be reduced to
the form
%%%%%
\bea
e_1\, B_{\u0\u2} + e_2\, B_{\u1\u2} + e_3\, N + e_4\,L +
e_5\, X+ e_6\, \Phi + e_7\, \Psi=0\,,\label{structures}
\eea
%%%%%
where each of the coefficients $e_i$ is given in terms of the $f_i$ and
their derivatives, together with quantities built from the 
background metric functions, etc.    Thus eqn (\ref{Zeqn}) is reduced to
the set of equations 
%%%%%
\bea
e_i=0\,,\qquad 1\le i\le 7.
\eea
%%%%%

   The simplest of the $e_i$ coefficients is $e_1$ which, after using 
eqns (\ref{bhsol}) and (\ref{backgroundrels}), becomes:
%%%%%
\bea
e_1 &=& 2(f_+\, f_-)^2\, f_-^{-\fft{4a^2}{1+a^2}}\, \Big[
f_1' - \Big(\fft1{r} + \fft{a^2\, \rms}{(1+a^2)\, r^2\, f_-}\Big)\, f_1 \nn\\
&&\qquad\qquad\qquad\quad +
f_3' - \Big(\fft2{r} + \fft{a^2\, \rms}{(1+a^2)\, r^2\, f_-}\Big)\, f_3\nn\\
&& \qquad\qquad\qquad\quad +
 \Big(\fft{2 a \,\rms}{(1+a^2)\, r^3\, f_-} - 
   \fft{2 a (n+1)}{r^2\, f_-\, f_+}\Big)\, f_5\Big]\,.\label{B02eqn}
\eea
%%%%%
Guided by what happened in the Reissner-Nordstr\"om case in
\cite{xanth2,chandra}, we choose to solve this
be requiring first that the terms involving $f_1$ should vanish, and the
terms involving $f_3$ should vanish, so
%%%%%
\bea
f_1' - \Big(\fft1{r} + \fft{a^2\,\rms}{(1+a^2)\, r^2\, f_-}\Big)\, f_1= 0\,,
\qquad f_3' -\Big(\fft2{r} + \fft{a^2\, \rms}{(1+a^2)\, r^2\, f_-}\Big)\, 
        f_3  =0\,.
\eea
%%%%%%
Thus we obtain
%%%%%
\bea
f_1 = c_1\, r\, f_-^{\fft{a^2}{1+a^2}}\,,\qquad
f_3= c_3\, r^2\, f_-^{\fft{a^2}{1+a^2}}\,,
\eea
%%%%%
where $c_1$ and $c_3$ are constants, as yet arbitrary.  Eqn 
(\ref{B02eqn}) now implies
%%%%%
\bea
f_5=0\,.
\eea
%%%%%

   The coefficient $e_7$ now becomes a simple first-order equation for $f_4$,
for which the solution to $e_7=0$ gives
%%%%%
\bea
f_4 = c_4\, r\, f_-^{\fft{a^2}{1+a^2}}\,,
\eea
%%%%%
where $c_4$ is another constant.
The condition that the coefficient $e_3$ vanishes now gives an algebraic
solution for $f_2$, namely
%%%%%
\bea
f_2 = -\fft{2 r\,K}{\Xi}\,  f_-^{\fft{a^2}{1+a^2}}\,,\label{solf2}
\eea
%%%%%
where we have defined
%%%%%
\bea
K(r) &=& n\,(1+a^2)^2\,c_1\, r^3 - (1+a^2)\,  
   \big[n\, (1+a^2)\, c_1 + 2 c_3\, \rps + 2 a\, c_4\big]\,\rms\, r^2 
\label{defF}\\
&&+ \big[2 (1+a^2)\, c_3\,\rms\, \rps + a\,(3-a^2)\,c_4\, \rms +
   3 a \, (1+a^2)\,c_4\, \rps\big]\,\rms\, r - 
               4a \, c_4\, \rps\, \rms^2\,,\nn
\eea
%%%%%%
and
%%%%%%
\bea
\Xi(r)&\equiv & 2n\, (1+a^2)^2\,r^3 + (1+a^2)\,
\big[3 (1-a^2)\, \rms + 3(1+a^2)\, \rps -
               2n\, (1+a^2)\, \rms\big]\, r^2 \nn\\
&& - \big[(3-a^2)\, \rms + 7 (1+a^2)\, \rps\big]\,\rms\, r +
4 \rps\, \rms^2\,.\label{defXi}
\eea
%%%%%
By then requiring that $e_6=0$ we can solve algebraically for the function $g$. 
The resulting expression for $g$ is rather complicated, and will require some
manipulation in order to make it presentable.  For now, we remark that
if the solution for $f_2$ is not yet substituted into the expression
for the coefficient $e_6$, then $g$ can be written manageably as
%%%%%
\bea
g&=& \beta_1\, f_2' + \beta_2\, f_2 + \beta_3\,,\nn\\
\beta_1 &=& -\fft{a\, \rms}{(1+a^2)\, c_4\, r^3}\, f_+^2\, 
f_-^{\fft{1-4a^2}{1+a^2}}\,,\nn\\
\beta_2 &=&\fft{a\, \rms}{(1+a^2)^2\, c_4\, r^6}\, f_+\, 
 f_-^{-\fft{5a^2}{1+a^2}}\, \Big[ (1+a^2)\, r\, (3r-4\rps) -3 r\, \rms +
   (4+a^2)\,\rms\, \rps\Big]\,,\nn\\
\beta_3&=& \fft{1}{(1+a^2)^2\, c_4\, r^5}\, f_+\, f_-^{-\fft{4a^2}{1+a^2}}\,
\Big\{2(n+1)\, (1+a^2)^2\,c_4\, r^3 -\nn\\
&&\qquad (1+a^2)\, \Big[2 a\,c_3\, \rms\, \rps +
 c_4\, \big[(1+3a^2)\, \rms -(1+a^2)\, \rps + 2n\, (1+a^2)\,\rms\big]
\Big]\, r^2\nn\\
&&\qquad
+\Big[ 2a(1+a^2)\, c_3\, \rms\, \rps + 
  c_4\, \big[(4a^2-1)\, \rms +(1+a^2)(2a^2-3)\, \rps\big]\Big]\, \rms\, r\nn\\
&&\qquad -(2a^4+5a^2-2)\, c_4\, \rms^2\, \rps
  \Big\}\,.\label{gres}
\eea
%%%%%

  From this point onwards, we are going to have to hope for miracles.
With the solutions we have now obtained for the functions 
$(f_1,f_2,f_3, f_4, f_5, g)$, we are left with just three constants, 
namely $(c_1,c_3,c_4)$, to play with.  And yet we still have to require
three more conditions to hold, namely $e_2=0$, $e_4=0$ and $e_5=0$.  These
are all equations that depend upon $r$, and so in order to 
satisfy the vanishing of all three functions, it must be that each
of the three functions $e_2$, $e_4$ and $e_5$ 
can be arranged to vanish simultaneously, for all $r$, for
some appropriate choice of the constants $c_1$, $c_3$ and $c_4$.
Putting each of the equations $e_2=0$, $e_4=0$ and $e_5=0$ over a 
common denominator and looking just at
the numerator, we obtain in each case a 
factor that is of the form of a polynomial in $r$ with coefficients
that are expressed in terms of the black hole parameters and the three
constants $c_1$, $c_3$ and $c_4$.  The coefficients of each power of
$r$ must separately vanish, yielding many algebraic equations. 

   We shall choose first the simplest of the equations from $e_4=0$ and the
simplest of the equations from $e_5=0$.  These two equations are each quadratic 
multinomials in the quantities $c_1$, $c_3$ and $c_4$.  Between these two
equations we can eliminate the quadratic powers of $c_4$ and hence obtain
an equation linear in $c_4$, which can be solved so as to express $c_4$ in
terms of $c_1$ and $c_3$:
%%%%%
\bea
c_4 = \fft{a\, c_1\, c_3\, \rms}{2[(1+a^2)\, c_1 + c_3\, \rms]}\,.
\label{solc4}
\eea
%%%%%%
It is convenient to introduce a dimensionless constant $q$ in place of
$c_1$, by defining
%%%%%
\bea
c_1= \fft{q\, c_3\, \rms}{2n\,(1+a^2)}\,,\label{c1q}
\eea
%%%%%
and so eqn (\ref{solc4}) then becomes
%%%%%
\bea
c_4=\fft{a\, q\, c_3\, \rms}{2(1+a^2)\, (q+2n)}\,.\label{c4q}
\eea
%%%%
Now, feeding eqns (\ref{c1q}) and (\ref{c4q})
back into either one of the two equations we selected above, we
obtain a cubic polynomial in $q$, which must vanish.  It is given by
%%%%%
\bea
\Lambda(q) &\equiv& q^3\, \rms +
 q^2\, \big[2n\,(1+a^2)\, \rms - 3(1-a^2)\, \rms - 3(1+a^2)\,\rps\big]
\nn\\
&& -2n\, q\,\big[(3-a^2)\,\rms +7(1+a^2)\,\rps\big]
-16n^2\, (1+a^2)\, \rps=0\,.\label{cubic}
\eea
%%%%%

   There is now no further freedom to choose any parameters.  
We can verify that, remarkably,  
$e_2$, $e_4$ and $e_5$ in fact vanish,
provided that (\ref{solc4}) and (\ref{cubic}) are satisfied.

   In summary, we have found that by making the ansatz (\ref{Zans}) and
plugging into eqn (\ref{Zeqn}), and then solving for the $f_i$ and $g$ 
functions as described above, we obtain three independent eigenfunctions
$Z^+$ that obey (\ref{Zeqn}), corresponding to the three different 
roots of eqn (\ref{cubic}).  We shall denote these eigenfunctions by
$Z^+_i$   In other words, we have succeeded in 
deriving from the system of seven first-order polar equations 
(\ref{7firstorder}) a
set of three decoupled second-order
differential equations.  The hoped-for miracles have occurred.  It is not
obvious why.

\bigskip
\noindent
{\bf Some further remarks}:

  It can be shown that the discriminant $D$ of the 
cubic equation (\ref{cubic}) is always non-negative 
(whenever $\rps\ge \rms$), and hence that
the three roots are always real.
This can be seen by writing $\rps =\rms\, (1+x)$ (so we always have $x\ge 0$),
and then $D$ can be written as $D= 4n^2\, \rms^4\, \wtd D$, where
%%%%%
\bea
\wtd D&=& (2+2n-x)^2\, \big[9x^2 + 4(8n+9)(x+1)\big] +
  a^8\, (2n-3x)^2 \big[x^2 + 4(8n+9)(x+1)\big] \nn\\
&&+ a^2\, S_1 + a^4\, S_2 +
  a^6\, S_3\,,
\eea
%%%%%
with $S_1$, $S_2$ and $S_3$ being certain multinomials in $x$ and $n$ for
which every term has a positive coefficient.  Consequently, it follows
that the three roots of the cubic are always real.\footnote{To be precise,
this argument proves that the discriminant is non-negative for all $x\ge0$
and $n\ge0$.  Since $n=\ft12\ell(\ell+1) -1$, the case $\ell=0$, for which 
$n=-1$, is special.  Setting $n=-1$, it can be seen that $\wtd D$ can then be 
written as $16a^4\,(1-a^2)^2 + S_4$, where $S_4$ is a multinomial in
$a$ and $x$ where all coefficients are manifestly positive.  So the
discriminant is non-negative also for $n=-1$.}
 
It is interesting to note that the cubic polynomial $\Lambda(q)$ defined
in (\ref{cubic}) and the cubic
function $\Xi(r)$ defined in eqn (\ref{defXi}) are related, in the
following sense:  
%%%%%
\bea
\Lambda(q) = -\fft{4n^2\, (1+a^2)}{\rms^2}\, 
  \Xi(r)\Big|_{r=\fft{-q\, \rms}{2n\, (1+a^2)}}=  
-\fft{4n^2\, (1+a^2)}{\rms^2}\,
  \Xi\Big(\!-\fft{c_1}{c_3}\Big)\,.
\eea
%%%%%

\section{Constructing The General Solution For Polar Modes}
\label{sec:polarmodes}

   In the analysis of the Reissner-Nordstr\"om perturbations by 
Chandrasekhar and Xanthopoulos \cite{xanth2,chandra}, 
two intermediate variables $H_1^+$ and
$H_2^+$ were introduced, as solutions of a coupled pair of second-order
differential equations.  Eventually, from these, linear combinations 
$Z^+_1$ and $Z^+_2$ were found, which obeyed decoupled second-order
equations.  In our discussion of the polar perturbations for
EMD black holes so far, we have essentially approached this
aspect of the problem from the other end, by first directly 
finding the decoupled
combinations $Z^+$ defined in eqn (\ref{Zans}) that satisfy the 
second-order equation in (\ref{Zeqn}).  There were three independent such
functions $Z^+_i$, corresponding to the three roots of the cubic 
equation (\ref{cubic}).  It will now prove to be advantageous to 
``reverse engineer'' from our results for our three $Z^+_i$, in order to
identify three functions $H^+_i$ from which the three $Z^+_i$ can
be obtained as certain linear combinations with constant coefficients.
The 3-vector $(H^+_1,H^+_2,H^+_3)^T$ will obey a matrix-valued second-order
differential equation that has the great advantage that it does 
not itself involve the complications inherent in manipulating the
roots of the cubic equation.  (These roots essentially then arise when 
one diagonalises the $3\times 3$ matrix characterising the
potential terms, to find the decoupled combinations $Z^+_i$.)
Apart from the considerably greater complexity of our larger system
of fields in the EMD case, the structure of the manipulations we shall 
need to perform will be qualitatively closely analogous to those carried
out for Reissner-Nordstr\"om in \cite{xanth2,chandra}.

\subsection{Introduction of the three $H^+_i$ eigenfunctions} 

  If we look at the two $H_i^+$ functions in \cite{xanth2,chandra}, we
see that $H_1^+$ is built as a linear combination, with fairly simple
$r$ dependent coefficients, of the variable $B_{\u1\u2}$ (translating to
our notation) and the variable $(L+X-B_{\u1\u2})$.  On the other hand,
their $H_2^+$ is built from a simple linear combination involving the function
$X$ and the function $(L+X-B_{\u1\u2})$.  Looking at our expression 
(\ref{Zans}) for the ansatz for $Z^+$ in our case, it is rather suggestive 
that, before solving for the constants $c_1$, $c_3$ and $c_4$, we
can decompose $Z^+$ as the sum of three terms, one having $c_1$ as coefficient,
one having $c_3$ as coefficient and one having $c_4$ as coefficient. 
Thus we have
%%%%%
\bea
Z^+ = c_3\, \hat H^+_1   + c_1\, \hat H^+_2 + c_4\, \hat H^+_3\,,\label{ZH0}
\eea
%%%%%
with
%%%%%
\bea
\hat H^+_1 &=& r\, e^{\bmu_2}\, B_{\u1\u2} + 
\fft{4 (1+a^2)\, r^2\, f_-\, \rms\, \rps}{\Xi}\,   e^{\bmu_2}\, 
  (L+X-B_{\u1\u2})\,,\nn\\
\hat H^+_2 &=& e^{\bmu_2}\, X - \fft{2n(1+a^2)^2\, r^3\, f_-}{\Xi}\, e^{\bmu_2}\, 
  (L+X-B_{\u1\u2})\,,\label{hatHi}\\
\hat H^+_3 &=& e^{\bmu_2}\, \Phi  + 
\fft{2 a \, \rms\, \Sigma}{\Xi}\, e^{\bmu_2}\,(L+X-B_{\u1\u2})\,,\nn
\eea
%%%%%
where $\Xi$ is defined in eqn (\ref{defXi}), 
%%%%%
\bea
\Sigma= 2(1+a^2)\, r^2 -\big(3(1+a^2)\, \rps + (3-a^2)\, \rms\big)\, r +
     4 \rms\, \rps\,,\label{defSigma}
\eea
%%%%%
and we used the expression (\ref{solf2}) for the function $f_2$.

\subsection{Finding the general solution}\label{sec:genpolar}

  At this stage, we have succeeded in reducing the system of seven first-order
equations (\ref{7firstorder}) to three second-order equations.  
However, since the latter are of total order
six, this means that the degree of freedom corresponding to the 
seventh order in the original first-order 
system remains to be isolated.  To put it
another way, ``solving the system'' requires that we obtain general 
expressions for the seven fields $(N,L,X,B_{\u1\u2},B_{\u0\u2},\Phi,\Psi)$
that satisfy the equations (\ref{7firstorder}), and not merely that we
establish that these equations {\it imply} the three second-order 
equations (\ref{secondorderop}).  Completing this construction requires
a fairly intricate sequence of steps, based on the work described in 
\cite{xanth1,xanth2,wald}, and it is to this that we now turn.

  First, we rescale the fields $\hat H^+_i$ defined in eqns (\ref{hatHi})
and define
%%%%%
\bea
H^+_1= b_1\, \hat H_1\,,\qquad H^+_2= b_2\, \hat H^+_2\,,
\qquad H_3=b_3\,\hat H_3\,,
\label{Hrescal}
\eea
%%%%%%
where we shall choose the constants $b_i$ so that the $H_i$ obey
%%%%%
\bea
(\del_{r_*}^2 +\omega^2)\, \begin{pmatrix} H^+_1\\ H^+_2\\ H^+_3\end{pmatrix} 
={\bf M}\, \begin{pmatrix} H^+_1\\ H^+_2\\ H^+_3\end{pmatrix}
\label{secondorderop}
\eea
%%%%%
with the $3\times 3$ matrix ${\bf M}$ being symmetric, and
the rescaled $H^+_1$ and $H^+_2$ reducing to those of \cite{xanth2} when
the dilaton coupling $a$ is set to zero.  This can be achieved by taking
%%%%%
\bea
b_1=-\fft{1}{Q\,\mu}\,,\qquad
b_2= \fft{1}{n}\,,\qquad b_3= \fft1{2\sqrt{n(n+1)}}\,.\label{kchoice}
\eea
%%%%%
(It is, seemingly, a quite non-trivial fact that the three ratios 
of transposition-related components of ${\bf M}$ are constants and that
all three ratios can be set to 1 by choosing the two non-trivial ratios of
the $b_i$ constants appropriately.)  
Thus with the choices for the $b_i$ as in eqns (\ref{kchoice}) then from
eqns (\ref{hatHi}) and (\ref{Hrescal}) we have
%%%%%
\bea
H^+_1 &=& -\fft{r}{Q\,\mu}\, e^{\bmu_2}\, B_{\u1\u2} -
\fft{4 (1+a^2)\, r^2\, f_-\, \rms\, \rps}{Q\, \mu\,\Xi}\, e^{\bmu_2}\,
  (L+X-B_{\u1\u2})\,,\nn\\
H^+_2 &=& \fft1{n}\, e^{\bmu_2}\, X - 
   \fft{2(1+a^2)^2\, r^3\, f_-}{\Xi}\, e^{\bmu_2}\,
  (L+X-B_{\u1\u2})\,,\label{finalHi}\\
H^+_3 &=& \fft1{2\sqrt{n(n+1)}}\, e^{\bmu_2}\, \Phi  +
\fft{a \, \rms\, \Sigma}{\sqrt{n(n+1)}\, \Xi}\, 
    e^{\bmu_2}\,(L+X-B_{\u1\u2})\,,\nn
\eea
%%%%%
(recall again that we are assuming $\ell\ge2$, and so $n$ is positive and 
$\mu$ is real and
positive) and the components of the $3\times 3$ matrix 
$\bM$ in eqn (\ref{secondorderop}) satisfy the symmetry property
%%%%%
\bea
\bM_{ij}=\bM_{ji}\,.
\eea
%%%%%
We see from the $a\rightarrow 0$ limit that $H^+_3$ represents a scalar
field degree of freedom.   

Since the explicit expressions for the components of the matrix $\bM$ are 
quite complicated, we relegate a presentation of $\bM$ to appendix A. 

\subsection{Treatment of the reducible first-order system}\label{sec:reducible}

   At this stage we have obtained the second-order equations 
satisfied by the three functions $H^+_i$.  In the notation of
\cite{xanth1}, and defining 
%%%%%
\bea
\bvphi= \begin{pmatrix} H^+_1 \\ H^+_2 \\ H^+_3 \end{pmatrix}\,,
\eea
%%%%%
these equations take the form 
%%%%%
\bea
\cO\, \bvphi=0\,,
\eea
%%%%%
where
%%%%%
\bea
\cO = \oneone\,\del_r^2 + \cB\, \del_r + \cC\label{cOdef}
\eea
%%%%%
and, comparing with eqn (\ref{secondorderop}), we therefore have
%%%%%
\bea
\cB= 2\bnu'\, \oneone\,,\qquad \cC= e^{-4\bnu}\, 
\big(\omega^2\, \oneone - \bM\big)\,.\label{cOterms}
\eea
%%%%%
Note that as in the general discussion in \cite{xanth1}, the second-order
operator $\cO$ is not self-adjoint.  

   Next, following \cite{xanth1}, we may read off from eqns (\ref{finalHi})
the $3\times 7$ matrix 
$\cL$ that implements the mapping 
%%%%%
\bea
\bvphi= \cL\, \cx\,,\qquad \hbox{where}\quad 
\cx=(N,L,X,B_{\u1\u2},B_{\u0\u2}, \Phi,\Psi)^T\,,
\eea
%%%%%
where the 7-vector $\cx$ is comprised of the seven variables that obey the
first-order equations given by (\ref{7firstorder}) and 
(\ref{abctilde}).  Thus we find
 %%%%%
\bea
\cL =\begin{pmatrix} 0& h_1 & h_1& h_2 &0&0&0\\
                     0& h_3 & h_4 & -h_3 &0 &0 &0\\
                     0& h_5 & h_5 & -h_5 & 0& h_6 & 0
                     \end{pmatrix}\,,\label{cLmat}
\eea
%%%%%
where 
%%%%%
\bea
h_1 &=& -\fft{4 (1+a^2)^2\,Q\, r^2\, e^{\bmu_2} \,f_-}{\mu\, \Xi} \,,\qquad
h_2= -h_1 -  \fft{r\, e^{\bmu_2}}{Q\, \mu}\,,\qquad
h_3 = -\fft{2 (1+a^2)^2\, r^3\, e^{\bmu_2}\, f_-}{\Xi}  \,,\nn\\
h_4 &=& h_3 + \fft{e^{\bmu_2}}{n}\,,\qquad
h_5 = \fft{a \, \Sigma \,\rms\,  e^{\bmu_2}}{\sqrt{n(n+1}\, \Xi}\,,\qquad
h_6 = \fft{e^{\bmu_2}}{2\sqrt{n(n+1)}}\,,
\eea
%%%%%   
and where $\Xi$ is defined in eqn (\ref{defXi}) and
$\Sigma$ is defined in eqn (\ref{defSigma}).

   Next, we write the seven first-order equations as
%%%%%
\bea
\cE \, \cx\equiv  \del_r \,\cx - \cA\,\cx=0\,,\label{cEdef}
\eea
%%%%%
where the $7\times 7$ matrix $\cA$ can be read off from the
first-order equations given in (\ref{7firstorder}) and (\ref{abctilde}):
%%%%%
\bea
\cA=\begin{pmatrix} \td a& \td b& \td c& -\td c& 0  &
  \gamma_1+\ft14\bar\phi'& \fft{\bar\phi'}{4\bmu_2} \\
(\td a-\bmu_2'+\bnu') & (\td b -\bmu_2'-\bnu') & \td c & -(\td c + \fft2{r}) &
    0 & \gamma_1+\ft34\bar\phi'&
     \fft{\bar\phi'}{4\bmu_2} \\
- (\td a-\bmu_2'+\bnu')& -(\td b +\bmu_2' -2\bnu') & -(\td c+\bmu_2'-\bnu') &
(\td c+\bmu_2'-\bnu') & 1 & -\gamma_1-\ft12\bar\phi'& 
                               -\fft{\bar\phi'}{4\bmu_2}\\
0& 0& 0& -\fft2{r} & 1 & 0& 0\\
\gamma_2 & 3 \gamma_2 & 2\gamma_2& \gamma_3& \gamma_4 & -a\,\gamma_2 &0\\
0&0&0&0&0&0&1\\
(\bnu'-\bmu_2')\,\bar\phi' & \gamma_5 & \gamma_6 & \gamma_7& 0 & 
 \gamma_8& \gamma_4
\end{pmatrix}\,,\nn\\
\label{cAdef}
\eea
%%%%%
with
%%%%%
\bea
\gamma_1&=& \fft1{4\bmu_2'}\,\Big(\bnu'\,\bar\phi' +
   \fft{2a\, Q^2}{r^2}\, e^{-2\bmu_2-2\bnu}\Big)\,,\qquad
\gamma_2=  \fft{2Q^2}{r^2}\, e^{-2\bmu_2-2\bnu}\,,\nn\\
\gamma_3&=& 2(n+1)\,e^{-2\bmu_2 -2\bnu}  
  -\omega^2\, e^{-4\bnu} -2\gamma_2\,,\qquad
\gamma_4 = -2(\bmu_2'+\bnu')\,,\nn\\
\gamma_5&=& (\bnu'-5\bmu_2')\, \bar\phi' -6 a\,\gamma_2\,,\qquad
\gamma_6=  2(\bnu'- \bmu_2')\,\bar\phi' -4a\, \gamma_2\,,\nn\\
\gamma_7&=& -\gamma_6 -\fft{2\bar\phi'}{r} 
  - 4a\,(n+1) \, e^{-2\bmu_2-2\bnu}\,,\nn\\
\gamma_8 &=& {\bar\phi{}'}^2 + 2(n+1)\, e^{-2\bmu_2-2\bnu} -
         \omega^2\,e^{-4\bnu} + a^2\, \gamma_2\,.\label{gammadef}
\eea
%%%%%

  Having found the matrices $\cL$ and $\cA$ we can then construct the
$3\times 7$ first-order matrix operator $\cS$ that has the property
%%%%%
\bea
\cS\, \cE = \cO\, \cL\,.\label{SE=OL}
\eea
%%%%%
As shown in \cite{xanth1}, it is given by
%%%%%
\bea
\cS= \cL\, \del_r + \cL' + \Gamma + \cB\, \cL\,,\qquad\hbox{where}\quad
\Gamma= \cL' + \cL\, \cA\,.
\eea
%%%%%
As can be seen by substituting this into eqn (\ref{SE=OL}), this
operator equation will be satisfied if
%%%%%
\bea
\cL'' + \cB\, \cL' + \cC\, \cL + 2\cL'\,\cA + \cL\, \cA' + \cL\, \cA^2 + 
\cB\, \cL\, \cA=0\,.\label{LABC}
\eea
%%%%%%
A straightforward calculation using our expressions above confirms that
(\ref{LABC}) is indeed satisfied.

   The $7\times 3$ adjoint matrix operator $\cS^\dagger$ is given 
by \cite{xanth1}
%%%%%
\bea
\cS^\dagger = -\cL^\dagger\, \del_r + {\cL'}^\dagger +
\cA^\dagger\, \cL^\dagger + \cB^\dagger\, \cL^\dagger\,.\label{Sdagdef}
\eea
%%%%%
Note that the matrix $\cB$ is proportional to the identity, 
with $\cB=2\bnu'\, \oneone$.
Following \cite{xanth1}, the 
operator $\cS^\dagger$ is then employed in order to construct 
%%%%%
\bea
\td \cx = 
  \cS^\dagger\, \bar\bvphi\,,\label{Sdageqn}
\eea
%%%%%
where we introduce
%%%%%
\bea
\bar\bvphi = (\bar H_1,\bar H_2,\bar H_3)^T\,,\qquad
 \td\cx=(\wtd N, \wtd L,\wtd X,\wtd B_{\u1\u2}, \wtd B_{\u0\u2},
\wtd \Phi,\wtd\psi)^T\,.\label{txbphi}
\eea
%%%%%

     The strategy then \cite{xanth1} is to
eliminate the three $\bar H_i$ and the three $\bar H_i'$ quantities
algebraically from the set of seven equations contained in
(\ref{Sdageqn}), thereby arriving at a constraint 
analogous to the relation $\Delta=0$ of eqn (29) in
\cite{xanth2}.  In other words, we can pick six out of the seven 
equations in (\ref{Sdageqn}) and solve them algebraically for the
three $\bar H_i$ and the three $\bar H_i'$ quantities, and then substitute
these solutions into the seventh equation to obtain the constraint relation.
In practice, we chose the equations from the components
$(1,2,3,4,5,7)$ of eqns (\ref{Sdageqn}) in order to
solve algebraically for the $\bar H_i$ and $\bar H_i'$, and we then used
these solutions to eliminate the $\bar H_i$ and $\bar H_i'$ from the
remaining component of (\ref{Sdageqn}) (\ie the unused 6th component), 
thus arriving
after some algebra at the constraint $\Delta=0$, with
%%%%%
\bea 
\Delta &=& n\, r\, \wtd X -\fft{2\rms\,\rps}{(1+a^2)\, r}\, \wtd B_{\u1\u2}
+ \fft{\rms\,\rps\, \Sigma}{(1+a^2)^2\, r^2}\, e^{-2\bmu_2-2\bnu}\,
  \wtd B_{\u0\u2} 
 -\fft{a\, \rms\, \Sigma}{(1+a^2)^2\, r\, (r-\rms)}\, \wtd\Phi \nn\\
&& + 
\Big[\fft{\Xi}{2(1+a^2)^2\, r\, (r-\rms)}- n\, r 
    -\fft{2 \rms\,\rps}{(1+a^2)\, r}\Big]\, \wtd L\nn\\
&&+\Big[-\omega^2\, r\, e^{2\bmu_2-2\bnu} +
\fft{\big[\big((1+a^2)\, \rps+(1-a^2)\, \rms\big)\, r  
  -2\rms\, \rps\big]\,\Sigma}{4(1+a^2)^2\, r}\, e^{-2\bnu-2\bmu_2}
     \Big]\, \wtd N\nn\\
&&+\Big[\fft{a\, \rms}{2(1+a^2)\, (r-\rps)}\nn\\
&&
\quad+
\fft{a\,\rms}{2(1+a^2)^3\, r^2\, (r-\rms)^2}\, 
\Big([7(1+a^2)^2\, r^3 -(1+a^2)(21(1+a^2)\, \rps +
  2(11-5a^2)\,\rms)\, r^2 \nn\\
&&\qquad +2\rms\,(3(1+a^2)(9+a^2)\, \rps
  +(7-2a^2-a^4)\,\rms)\, r -16(2+a^2)\, \rms^2\, \rps \Big)\Big]\, \wtd\Psi\,.
\label{Deltares}
\eea
%%%%%
(We are following the notation of \cite{xanth2,chandra} in using the
symbol $\Delta$ for this constraint, even though $\Delta$ was also used
previously for the completely unrelated function appearing in the
background
metric.)  Note that we have made a choice for 
the arbitrary overall scale function in
this definition of $\Delta$ such that the coefficient of $\wtd X$
coincides with that for the Reissner-Nordstr\"om case in \cite{xanth2}
when $a=0$. 
It is useful also to record that the expressions for $\bar H_1$, $\bar H_2$ and
$\bar H_3$ obtained in the procedure described above are
%%%%%
\bea
\bar H_1 &=& -Q\,\mu\,\Big( \fft1{r}\, e^{-\bmu_2}\, \wtd B_{\u0\u2} +
   \fft{2(1+a^2)\,\big[(1+a^2)\, r - \rms\big]}{\Xi}\, 
  e^{\bmu_2+2\bnu}\,\wtd N\Big)\,,\nn\\
\bar H_2 &=& - \fft{2(1+a^2)\, n\, r\, \big[(1+a^2)\, r -\rms\big]}{
   \Xi}\, \wtd N\,,\nn\\
\bar H_3 &=& 2\sqrt{n\,(n+1)}\, e^{-\bmu_2}\, 
\Big(\wtd\Psi+ \fft{a\, r\, (r-\rms)(r-\rps)}{\Xi}\, \wtd N \Big)\,.
\label{barHdefs}
\eea
%%%%%

   We now calculate $\Delta'/\Delta$, making use of the adjoint 
first-order equations \cite{xanth1}
%%%%%
\bea
(\wtd N',\wtd L',\wtd X',\wtd B_{\u1\u2}',\wtd B_{\u0\u2}',
\wtd\Phi',\wtd\Psi')^T=
-\cA^\dagger\,(\wtd N,\wtd L,\wtd X,\wtd B_{\u1\u2},
          \wtd B_{\u0\u2},\wtd\Phi,\wtd\Psi)^T\,,\label{tildefo}
\eea
%%%%%%
\ie by using the equations $\cE^\dagger\, \tilde\cx=0$.
This gives
%%%%%
\bea
\fft{\Delta'}{\Delta} = \fft{4(1+a^2)\, r^2 -\big[5(1+a^2)\, \rps +
  (5+a^2)\, \rms\big]\, r 
   + 2(3+a^2)\, \rms\,\rps}{2(1+a^2)\, r(r-\rms)(r-\rps)}\,,
\eea
%%%%%
which may be integrated to give the simple result, re-expressed in terms
of the background metric functions, as
%%%%%
\bea
\Delta = r\, e^{\bmu_2-\bnu}\, \wtd\Delta\,,\label{tildeDelta}
\eea
%%%%%%
where $\wtd\Delta$ is an arbitrary constant of integration.

 By the general arguments in \cite{xanth1}, the coefficients of the
seven fields $(\wtd N,\wtd L,\wtd X,\wtd B_{\u1\u2},
\wtd B_{\u0\u2},\wtd\Phi,\wtd\Psi)$ in the expression for $\wtd\Delta=
 r^{-1}\, e^{\bnu-\bmu_2}\, \Delta$ with $\Delta$ given by eqn 
(\ref{Deltares}) are a set of functions
$(N_\0,L_\0,X_\0,B_{12\0},B_{02\0},\Phi_\0,\Psi_\0)$ that provide
a particular solution of the original first-order system of equations
(\ref{7firstorder}). Thus we have
%%%%%
\bea
N_\0 &=&
-\omega^2\,  e^{\bmu_2-\bnu} +
\fft{\big[\big((1+a^2\, \rps + (1-a^2)\, \rms\big)\, r - 
          2 \rms\,\rps\big]\,\Sigma}{4(1+a^2)^2\, r^2}\, e^{-\bnu-3\bmu_2}\,,\nn\\
L_\0 &=& \Big[\fft{\Xi}{2(1+a^2)^2\, r^2\, (r-\rms)} -n- 
\fft{2\rms\,\rps}{(1+a^2)\, r^2}\Big]\,  e^{\bnu-\bmu_2}\,,\nn\\
X_\0 &=&  n\, e^{\bnu-\bmu_2}\,,\nn\\
B_{\u1\u2\0}&=& -\fft{2\rms\,\rps}{(1+a^2)\, r^2}\,  e^{\bnu-\bmu_2}\,,\nn\\
B_{\u0\u2\0}&=& 
 \fft{\rms\,\rps\, \Sigma}{(1+a^2)^2\, r^3}\, e^{-\bnu -3\bmu_2}\,,\nn\\
\Phi_\0&=& - 
\fft{a\, \rms\, \Sigma}{(1+a^2)^2\, r^2\, (r-\rms)}\,e^{\bnu-\bmu_2} \,,\nn\\
\Psi_\0 &=& r^{-1}\, e^{\bnu-\bmu_2}\,
\Big[ \fft{a\, \rms}{2(1+a^2)\, (r-\rps)} \nn\\
&&
\!\!\!\!\!\!\!\! +
\fft{a\,\rms}{2(1+a^2)^3\, r^2\, (r-\rms)^2}\, 
\Big([7(1+a^2)^2\, r^3 -(1+a^2)\big(21(1+a^2)\, \rps +
  2(11-5a^2)\,\rms\big)\, r^2 \nn\\
&&\qquad +2\rms\,\big(3(1+a^2)(9+a^2)\, \rps
  +(7-2a^2-a^4)\,\rms\big)\, r -16(2+a^2)\, \rms^2\, \rps \Big)\Big]\,,\nn\\
&=& \Phi_\0'\,,
\label{x0sol}
\eea
%%%%%
where $\Sigma$ is defined in eqn (\ref{defSigma}).
It is a straightforward exercise, with the aid of an algebraic computing
package, to verify directly that these expressions indeed
solve the first-order system of equations (\ref{7firstorder}).

  The next step is to act on the 3-vector
$\bar\bvphi= (\bar H_1,\bar H_2,\bar H_3)^T$ with the adjoint second-order
operator $\cO^\dagger$, which from eqn (\ref{cOdef}) together with
(\ref{cOterms}), is given by
%%%%%
\bea
\cO^\dagger = \oneone\,\del_r^2 - 2\bnu'\, \del_r -2\bnu''\, \oneone +
   e^{-4\bnu}\, (\omega^2\,\oneone -\bM)
\eea
%%%%%
(recalling that $\bM$ is real and symmetric).  After some considerable
algebra, and making use of 
the first-order equations (\ref{tildefo}), we find that
%%%%%
\bea
\cO^\dagger \, \begin{pmatrix} \bar H_1 \\ \bar H_2\\ \bar H_3\end{pmatrix}=
 \begin{pmatrix} v_1 \\ v_2 \\ v_3\end{pmatrix}\, \wtd\Delta\,,
\eea
%%%%%
where
%%%%%
\bea
v_1 &=& \fft{2(1+a^2)\, Q\, \mu\, \big[(1+a^2)\, r - \rms\big]}{\Xi}\, 
e^{-\bnu}\,,\nn\\
v_2 &=& \fft{2(1+a^2)\, n\, r\, \big[(1+a^2)\, r - \rms\big]}{\Xi}\, 
e^{-\bnu}\,,\nn\\
v_3 &=& -\fft{2a\, (1+a^2)\, \sqrt{n\,(n+1)}\,r\, \rms }{\Xi}\,  
e^{-\bnu}\,.\label{v1v2v3}
\eea
%%%%%

   Having obtained the $v_i$, we then, following \cite{xanth1}, introduce a 
3-vector $\boldsymbol\kappa=(k_1(r),k_2(r),k_3(r))^T$ defined to obey the 
equations
%%%%%
\bea
\cO^\dagger\, \begin{pmatrix} k_1 \\ k_2 \\ k_3\end{pmatrix} =
  - \begin{pmatrix} v_1 \\ v_2 \\ v_3\end{pmatrix}\,.\label{keoms}
\eea
%%%%%
We also introduce new functions $\wtd H_i$ defined by
%%%%%
\bea
\wtd H_i = \bar H_i + k_i\, \wtd\Delta\,,\label{tildeHdef}
\eea
%%%%%
where the functions $\bar H_i$ are given by eqns (\ref{barHdefs}).  As noted
in \cite{xanth1}, in our present case 
these reduce the adjoint system of seven first-order
equations (\ref{tildefo}) to the set of three second-order equations
%%%%%
\bea
\cO^\dagger\, \begin{pmatrix} \wtd H_1 \\ \wtd H_2 \\ 
                     \wtd H_3\end{pmatrix} =0\,.
\eea
%%%%%

     By analogy with the matrix $\cL$ that was introduced earlier to 
relate the three functions in $\bvphi=(H^+_1,H^+_2,H^+_3)^T$ to the
seven functions $\cx=(N,L,X,B_{\u1\u2},B_{\u0\u2},\Phi,\Psi)^T$ by writing 
$\bvphi=\cL\, \cx$, we now introduce the $3\times 7$ 
matrix $\wtd\cL$ such that
%%%%%
\bea
\begin{pmatrix} \wtd H_1 \\ \wtd H_2 \\ \wtd H_3\end{pmatrix}
=
\wtd\cL\, \begin{pmatrix} \wtd N\\ \wtd L \\ \wtd X\\ \wtd B_{\u1\u2}\\
  \wtd B_{\u0\u2}\\ \wtd\Phi \\\ \wtd\Psi \end{pmatrix}\,.
\eea
%%%%% 
The matrix $\wtd\cL$ can be read off from eqn (\ref{tildeHdef}),
together with eqns (\ref{barHdefs}) and (\ref{Deltares}), with 
eqn (\ref{tildeDelta}) providing the definition of $\wtd\Delta$ in terms
of $\Delta$.

  As was shown in \cite{xanth1}, one can now introduce a first-order
matrix operator $\wtd\cS$ having the property that 
%%%%%
\bea
\wtd\cS\,\cE^\dagger= \cO^\dagger\, \wtd\cL\,,\label{tildeSprop}
\eea
%%%%%
 where $\cE^\dagger = -\oneone\,\del_r -\cA^\dagger$
is the adjoint of the original 
first-order matrix operator $\cE$ given by eqns 
(\ref{cEdef})--(\ref{gammadef}).
The operator
$\wtd\cS$ and its adjoint are given by \cite{xanth1}
%%%%%
\bea
\wtd\cS &=& \wtd\cL\, \del_r + 2 \wtd\cL'  - \wtd\cL\, \cA^\dagger -
  \cB^\dagger\, \wtd\cL\,,\nn\\
\wtd\cS^\dagger &=& -\wtd\cL^\dagger \, \del_r + {\wtd\cL^\dagger}{'}  - 
  \cA\,\wtd\cL^\dagger -  \wtd\cL^\dagger\,\cB\,.
\eea
%%%%%
By taking the adjoint of eqn (\ref{tildeSprop}), it can be seen that
if we define the quantity $\cx$ by $\cx=\wtd\cS^\dagger\, \bvphi$ then 
%%%%%
\bea
\cE\,\cx= \cE\,\wtd S^\dagger\,\bvphi = \wtd\cL^\dagger\,\cO\,\bvphi\,,
\eea
%%%%
showing that this gives a solution of the original first-order system 
$\cE\,\cx=0$ for every 
set of fields $\bvphi$ that solves the second-order equations 
$\cO\, \bvphi=0$ \cite{xanth1,wald}, given in our case by 
eqns (\ref{secondorderop}).
The resulting expressions for the seven fields are
%%%%%
\bea
N &=& N_\0 \, F + \Big[\fft{2(n+1)\, (1+a^2)^2\, r}{\Xi\, (r-\rps)}
\, e^{2\bnu+\bmu_2} - e^{2\bnu}\, \del_r\, 
  \Big(\fft{2(1+a^2)}{\Xi}\, [(1+a^2)\, r - \rms]\, e^{\bmu_2}\Big)\Big]\, G
\nn\\
&&+ \fft{2(1+a^2)}{\Xi}\, [(1+a^2)\, r -\rms]\, e^{2\bnu+\bmu_2}\, \del_r\, G
 -\fft{4(1+a^2)\, n\, [(1+a^2)\, r - \rms]}{\Xi}\, e^{2\bnu+\bmu_2}\, H^+_2
\nn\\
&&+\fft{a\,\sqrt{n(n+1)}\,\rms}{[(1+a^2)\, r -\rms]}\,\Big[ e^{-\bmu_2} -
      \fft{4(1+a^2)\, \rms}{\Xi}\, e^{2\bnu+\bmu_2}\Big]\, H^+_3\,,\nn\\
L&=& L_\0\, F + 
   \fft{1}{r}\, e^{-\bmu_2}\, [Q\,\mu\, H^+_1 + n\, r\, H^+_2]\,,\nn\\
X&=& X_\0\, F - n\, e^{-\bmu_2}\, H^+_2\,,\nn\\
B_{\u1\u2}&=& B_{\u1\u2\0}\, F + \fft{Q\, \mu}{r}\, 
e^{-\bmu_2}\, H^+_1\,,\nn\\
B_{\u0\u2} &=& B_{\u0\u2\0}\, F + 
\fft{4 (1+a^2)\, Q^2\, [(1+a^2)\, r-\rms]}{r^2\,\Xi}\, e^{-\bmu_2}\,
  \Big[n\, r\, H^+_2 -\fft{a\, \sqrt{n(n+1)}\, r\, \rms}{[(1+a^2)\, r -\rms]}\,
  H^+_3\Big] \nn\\
&&+ \fft{\mu\, Q^3}{r^2}\, e^{-\bmu_2}\, 
\Big[\fft{4(1+a^2)\, [(1+a^2\, r - \rms]}{\Xi} - \fft{a^2}{(r-\rms)\, \rps}
   \Big]\, H^+_1 + \fft{\mu\, Q}{r}\, e^{-\bmu_2}\, H^+_1{'}\,,\nn\\
\Phi &=& \Phi_\0\, F - 2\sqrt{n(n+1)}\, e^{-\bmu_2}\,  H^+_3\,,\nn\\
\Psi &=& \Psi_\0\, F + 
\fft{2\sqrt{n(n+1)}\, [(1+a^2)\, r - \rms]}{(1+a^2)\, r\, (r-\rms)}\,
e^{-\bmu_2}\, H^+_3 - 2\sqrt{n(n+1)}\, e^{-\bmu_2}\, H^+_3{'}\label{finalsol}\\
&&+\fft{2a\, [(1+a^2)\, r - \rms]\, \rms}{
         (1+a^2)\, \Xi\, r^2\,(r-\rms)}\, e^{-\bmu_2}\, 
   \Sigma\, G\,,\nn
\eea
%%%%%
where $\Xi$ is defined in eqn (\ref{defXi}), $\Sigma$ is defined in 
eqn (\ref{defSigma}), and we have defined
%%%%%
\bea
G= Q\, \mu\, H^+_1 + n\, r\, H^+_2 -\fft{a\,\sqrt{n(n+1)}\, r\, \rms}{
                                 [(1+a^2)\, r - \rms]}\, H^+_3\,.\label{Gdef}
\eea
%%%%%
The functions $(N_\0,L_\0,X_\0,B_{\u1\u2\0},B_{\u0\u2\0},\Phi_\0,\Psi_\0)$ 
are those of the particular solution (\ref{x0sol}) obtained earlier, and
the function $F$ is given by
%%%%%
\bea
F=\sum_{i=1}^3 \Big( k_i'\, H^+_i - k_i\, H^+_i{'} - 
               2\bnu'\, k_i\, H^+_i\Big)\,.
\label{Fexp}
\eea
%%%%% 

It can be seen by differentiating eqn
(\ref{Fexp}), and making use of the of the second-order 
equations (\ref{secondorderop})
obeyed by the functions $H^+_i$, and the second-order equations (\ref{keoms})
obeyed by the functions $k_i$, that
%%%%%
\bea
F' = -\sum_{i=1}^3 v_i\, H^+_i\,,
\eea
%%%%%
where the $v_i$ are given by eqns (\ref{v1v2v3}), and so we have
%%%%%
\bea
F'= -\fft{2(1+a^2)\, [(1+a^2)\, r - \rms]}{\Xi}\, e^{-\bnu}\, G\,,
\eea
%%%%%
where $G$ is given in terms of the functions $H^+_i$ by eqn (\ref{Gdef}).
Thus $F$ can be expressed purely in terms of the three functions
$H^+_i$ that satisfy eqns (\ref{secondorderop}), as
%%%%%
\bea
F(r) &=& -\int^r \fft{2(1+a^2)\, [(1+a^2)\, r' - \rms]}{\Xi(r')}\, 
e^{-\bnu(r')}\, G(r')\, dr'\,,\label{Fint}\\
&=& -\int^r \fft{2(1+a^2)\, [(1+a^2)\, r' - \rms]}{\Xi(r')}\,
e^{-\bnu(r')}\, \Big[ Q\, \mu\, H^+_1(r') + n\, r'\, H^+_2(r')\nn\\
&&\qquad\qquad\qquad\qquad\qquad\qquad\qquad\qquad\qquad       -
\fft{a\,\sqrt{n(n+1)}\, r'\, \rms}{
                      [(1+a^2)\, r' - \rms]}\, H^+_3(r')\Big]\, dr'
\,.\nn 
\eea
%%%%%

    As a side remark, it 
can be verified by using eqns (\ref{finalsol}), the
function $F$ is also related to the fields $L$, $X$ and $B_{\u1\u2}$ by
$L+X-B_{\u1\u2}= (L_\0+X_\0-B_{\u1\u2\0})\, F$, and so, using eqns
(\ref{x0sol}), we have
%%%%%
\bea
F = \fft{2(1+a^2)^2\, r^2\, (r-\rms)}{\Xi}\, e^{-\bnu+\bmu_2}\, 
(L+X-B_{\u1\u2})\,.
\eea
%%%%%

  To summarise, we have established that the functions $(N,L,X,B_{\u1\u2},
B_{\u0\u2},\Phi,\Psi)$ given in eqns (\ref{finalsol}) solve the 
original set of seven 
first-order equations (\ref{7firstorder}).  These solutions are expressed in
terms of the three arbitrary solutions $(H^+_1,H^+_2,H^+_3)$ of the 
second-order equations (\ref{secondorderop}), together with one 
arbitrary constant 
multiple of the particular solution $(N_\0,L_\0,X_\0,B_{\u1\u2\0},
B_{\u0\u2\0},
\Phi_\0,\Psi_\0)$ obtained in eqns (\ref{x0sol}).  (The arbitrary constant
arises as the arbitrary additive parameter in the indefinite integral 
(\ref{Fint}).)  Thus we have established that the most general solution of 
the system of seven first-order equations (\ref{7firstorder}) is given in
terms of the three general solutions $(H^+_1,H^+_2,H^+_3)$ of the second-order
equations (\ref{secondorderop}), and one further constant of 
integration.\footnote{After some considerable algebra, it
can be verified explicitly that the expressions given in (\ref{finalsol})
satisfy the first-order equations (\ref{7firstorder}), provided that the
functions $H^+_i$ satisfy the second-order equations (\ref{secondorderop}).}

\section{Mode Stability And Positivity Of The Potentials}

  Having obtained the separated and diagonalised second-order equations for
the axial and the polar perturbations, we may now address the question of
mode 
stability of the EMD black holes solutions against linearised perturbations.
This comes down to establishing the positivity, outside the horizon, 
of the potentials in the 
second-order radial equations. First, we present a general discussion about
the notion of mode stability.

\subsection{The criterion for mode stability}

  We recall that in the mode analysis we are considering, the
modes are in fact {\it quasi-normal}; that is to say, their frequencies
are complex, $\omega=\omega_1 + \im \omega_2$.  Establishing mode stability
then amounts to showing that the imaginary parts $\omega_2$ of their
complex frequencies are always negative, since a mode with positive
$\omega_2$ would have a time dependence $e^{-\im\omega t}$ that grew 
exponentially at large $t$, thus implying an unstable perturbation.  
The boundary conditions on the modes are
that they should be {\it ingoing} on the future horizon and {\it outgoing}
near infinity, so the waves should be of the form\footnote{The 
radial and time dependence of the modes is governed by the second-order
equation (\ref{waveeqn}), and since the potential
$V$ goes to zero on the horizon and at infinity this implies the asymptotic
forms $Z\sim e^{-\im \omega t(t\pm r_*)}$ as seen in eqns (\ref{bcs}).}

%%%%%
\bea
\hbox{Near horizon}\,,\ r_*\longrightarrow -\infty:&& Z \sim 
e^{-\im\omega(t+ r_*)}\,,\nn\\
\hbox{Near infinity}\,,\ r_*\longrightarrow +\infty:&& Z\sim
  e^{-\im\omega(t-r_*)}\,.\label{bcs}
\eea
%%%%%
Note that if a mode with positive $\omega_2$ did exist, it would 
therefore fall off exponentially at both the horizon and infinity.

   The general form of the second-order equations for the separated 
and diagonalised
perturbations is $\big(\del_{r_*}^2 + \omega^2 - V\big) Z=0$, where $V$ is the
potential. Re-introducing the time dependence, we have
%%%%%
\bea
\del_{r_*}^2 Z - \del_t^2 Z - V\, Z=0\,.\label{waveeqn}
\eea
%%%%%
Following a general argument given in \cite{wald2}, we
multiply eqn (\ref{waveeqn})  
by $\del_t \bar Z$, and then add the complex conjugate of
this resulting equation. After a little manipulation, this can be seen
to give
%%%%%
\bea
\del_{r_*}\Big(\del_t\bar Z\, \del_{r_*} Z +\del_t Z\, \del_{r_*} \bar Z\Big)
-\del_t
\Big(|\del_{r_*} Z|^2 + |\del_t Z|^2 + V\, |Z|^2\Big)=0\,,\label{Zid1}
\eea
%%%%%
since the potential $V$ is independent of $t$.  Integrating from the
horizon to infinity using the $r_*$ variable, we therefore see that
%%%%%
\bea
\del_t \int_{-\infty}^\infty dr_*\, 
\Big(|\del_{r_*} Z|^2 + |\del_t Z|^2 + V\, |Z|^2\Big)=0\,,
\eea
%%%%%
provided we can assume that the boundary integral arising 
from the first term in eqn (\ref{Zid1})
vanishes.  The energy
%%%%%
\bea
E= \int_{-\infty}^\infty dr_*\, 
\Big(|\del_{r_*} Z|^2 + |\del_t Z|^2 + V\, |Z|^2\Big)\label{energy}
\eea
%%%%%
of the perturbation will then be constant in time.

  As we discussed above, if an unstable mode (\ie one with a positive imaginary
part $\omega_2$ in its frequency $\omega$) did exist, then it would
fall off exponentially rapidly as $r_*$ went to $-\infty$ (on the horizon) and 
as $r_*$ went to $+\infty$ (at infinity), and so for such a mode the
boundary integral coming from the first term in eqn (\ref{Zid1}) would indeed
vanish.   If the potential
$V$ is everywhere non-negative for $r_+\le r\le\infty$, it then follows that
the perturbation $Z$ would remain bounded since the energy $E$ in eqn 
(\ref{energy}) is conserved.  This would contradict
the supposed existence of the mode with positive $\omega_2$, since for
such a mode $E$ would grow exponentially in time, 
$E\sim E_0 \,e^{2\omega_2\, t}$.  Thus the possibility
of such modes is ruled out, establishing 
mode stability for this class of perturbations, provided that we can show
the positivity of the potential $V$ outside the horizon.

\subsection{Positivity of the axial-mode potentials}

      Writing out the expressions for the potentials 
$V^-_{1,2}$ in eqn (\ref{VM12}), it can
be seen that they can be written as a manifestly positive function times
$P_\pm$, where
%%%%%
\bea
P_\pm =P_0 \pm D\, P_1\,,\label{Ppm}
\eea
%%%%%
and where
%%%%%
\bea
P_0 &=& 
-(1+a^2)\, r^2 \big[(1+a^2) \,C  +2n\,  (1+a^2) \,\rms 
    +2 \rms \big]\nn\\
&&+r\, \Big[(1+a^2)^2\, C\, \rms  
  +(1+a^2)^2\, (4-3a^2)\, Q^2  -a^2\,(2-a^2)\, \rms^2   \Big]\nn\\
&&+2
    \left(a^2+1\right)^2 \,(n+1) r^3-\left(a^2+1\right) \left(a^2+4\right) 
\, Q^2\, \rms\,,\label{P0}\\
P_1 &=& (1+a^2)^2\, r\, (r-\rms)\,.\label{P1}
\eea
%%%%%
and
%%%%%
\bea
D= \sqrt{C^2+ 4 \mu^2\, Q^2}\,.\label{Ddef}
\eea
%%%%%
The expressions for $C$ and $\mu$ are given in eqn (\ref{Cmudef}).

Looking at eqn (\ref{VM12}), it is evident that the potential $V^-_{2}$,
corresponding the case $P_-$ in eqn (\ref{Ppm}), is the one that is 
more at risk of being negative.
For the potentials $V^-_1$ and $V^-_2$ both to be non-negative it would 
necessarily have
to be that $(V^-_1 + V^-_2)$ was non-negative, and so from eqn (\ref{VM12})
we see that we must first establish that $(-T)$ is non-negative.  Proving
this is equivalent to proving that the term $P_0$ in eqn (\ref{Ppm}) is 
non-negative.   From eqn (\ref{P0}), it is easy to see that by setting
%%%%%
\bea
\rps = \rms\, (1+x)\,,\qquad r=\rps\, (1+z)\,,\qquad n= 1+y\,,\label{xyzsub}
\eea
%%%%
the expression for $P_0$ is a manifestly positive prefactor times a 
66-term multinomial in $x$, $y$, $z$  and $a$ with all
positive coefficients, and hence $P_0$ is non-negative for all $r\ge \rps$
for all modes with $n\ge 1$ (\ie $\ell\ge 2$) for all the EMD black holes.

   Since $P_+=P_0+ D\, P_1$ and $P_1$ in eqn (\ref{P1}) is manifestly 
non-negative, it follows that $P_+$ is always non-negative outside the
horizon.  Establishing that $P_-$ is also non-negative is then equivalent
to establishing that $(P_0^2 - D^2\, P_1^2)$ is non-negative, and since
this does not involve square roots we can again make the substitutions in
eqn (\ref{xyzsub}), thereby obtaining a manifestly positive 
prefactor times a 480-term multinomial in $x$, $y$, $z$ and $a$.  It is easily 
seen that all the terms in the multinomial have
positive coefficients, and hence this shows that $P_-$ is always non-negative
outside the horizon for all $\ell\ge2$ modes, for all EMD black holes.

   This completes the demonstration that the axial-mode potentials $V^-_1$
and $V^-_2$ are always non-negative outside the horizon for all $\ell\ge 2$
modes and for all EMD black holes.  It is worth remarking also that the
asymptotic behaviours of the potentials $V^-_{1,2}$ near the horizon and
near infinity take the form
%%%%%
\bea
\hbox{For}\quad r \rightarrow \rps:&& V^-_{1,2} \sim (r-\rps) + 
        {\cal O}\Big((r-\rps)^2\Big)\,,\nn\\
\hbox{For}\quad r\rightarrow\infty:&& V^-_{1,2} = \fft{\ell(\ell+1)}{r^2} +
  {\cal O}\Big(\fft1{r^3}\Big)\,.\label{asymptotics}
\eea
%%%%%

\subsection{Positivity of the polar-mode potentials}

  For the polar perturbations, the relevant potentials are the functions
$g_i$ appearing in eqn (\ref{Zieqn}).  In principle we would wish
to study the potentials $g_i$ for arbitrary values of $a$, $\rms$, 
$\rps$ (subject to $\rps>\rms$), and $n=\ft12 (\ell+2)(\ell-1)$.  This,
however, would be an unwieldy procedure, since the specific form of the 
potential $g_i$ depends on the $i$'th root $q_i$ of the cubic polynomial
(\ref{cubic}).  It turns out to be advantageous instead to parameterise
the black hole solutions in terms of $(a,\rms,q,\ell)$ rather than
$(a,\rms,\rps,\ell)$ where we now instead solve the cubic equation 
(\ref{cubic}) for $\rps$ rather than $q$.  Eqn (\ref{cubic}) is
linear in $\rps$, and so this gives the simple result that
%%%%%
\bea
\rps= \fft{q\, \big[q^2 + q\, \big( 2n(1+a^2)+3(1-a^2)\big)
    -2n\, (3-a^2)\big]\, \rms}{(1+a^2)(q+2n)(3q + 8n)}\,.
\label{solrp}
\eea
%%%%% 
The quantity $q$ may now be viewed as a free parameter, subject only to
the requirement that we must have $r_+ > r_-$. This is in fact quite easy
to characterise, since it may be seen from eqn (\ref{solrp}) 
that after rewriting $n$ in terms of $\ell$ as in eqn (\ref{Xndef}), 
$(r_+-r_-)$ factorises to give
%%%%%
\bea
\rps - \rms = \fft{[q+ 2(\ell-1)]\, [q-2(\ell+2)]\,
    [q+ (1+a^2)(\ell+2)(\ell-1)]\, \rms}{(1+a^2)\, [q +(\ell+2)(\ell-1)]\,
      [3q +4(\ell+2)(\ell-1)]}\,.\label{rpsol}
\eea
%%%%  
The fact that this expression has factorised over the rationals is 
quite fortunate, and, as we shall see, it helps greatly with the 
following analysis.

  From eqn (\ref{rpsol}) it can be seen that there are 
three disjoint ranges of the parameter $q$ for 
which $(\rps - \rms)$ is positive;
the characterisations of the first two are independent of the value
of $a$, whilst the third is characterised differently depending on 
whether (3a): $a<\fft1{\sqrt3}$ or (3b): $a>\fft1{\sqrt3}$:
%%%%%
\bea
\hbox{\bf(1)}:&& q >2(\ell+2)\,,\nn\\
\hbox{\bf(2)}:&& -(\ell+2)(\ell-1) < q < -2(\ell-1)\,,\nn\\
\hbox{\bf(3a)}:&&  -\ft43 (\ell+2)(\ell-1) < q 
                       < -(1+a^2)(\ell+2)(\ell-1)\,,
 \qquad\hbox{or}\nn\\
\hbox{\bf(3b)}:&& -(1+a^2)(\ell+2)(\ell-1) <q < 
          -\ft43 (\ell+2)(\ell-1)
\,,\label{cases}
\eea
%%%%%
(The parameter range for $q$ contracts to zero for the cases (3a) or (3b) 
if $a=\fft1{\sqrt3}$.)  

  By parameterising the black holes in terms of $(a,\rms,q,\ell)$ as we have 
just described, we can avoid any need to work explicitly with the solutions
$q_i$ of the cubic equation (\ref{cubic}) for $q$.  After expressing
$\rps$ in terms of $q$ as in eqn (\ref{rpsol}), the expression for
the potential will be of the form $g=g(a,\rms,q,\ell,r)$, and this will be
a rational function of the arguments.  We first remark that near the horizon
and near infinity we have
%%%%%
\bea
\hbox{For}\quad r \rightarrow \rps:&& g\sim (r-\rps) + 
        {\cal O}\Big((r-\rps)^2\Big)\,,\nn\\
\hbox{For}\quad r\rightarrow\infty:&& g = \fft{\ell(\ell+1)}{r^2} +
  {\cal O}\Big(\fft1{r^3}\Big)\,,
\eea
%%%%%
which are of the same form as the asymptotic behaviours of the axial mode
potentials (\ref{asymptotics}).  
We now need to examine the sign of the potentials in the entire region
$\rps\le r\le \infty$.  

  Although the potential $g=g(a,\rms,q,\ell,r)$ is a rational function of
its arguments, it is quite a complicated one, which makes it difficult to
present an explicit demonstration of its positivity.  However, by good
fortune it turns out that the following analytic 
procedure allows us to establish its positivity, even though the steps are
only realistically implementable by means of algebraic computing techniques.
There are three parameter ranges for $q$ that need to be considered,
namely the intervals (1), (2) and (3a) (if $a<\fft1{\sqrt3}$); or
(1), (2) and (3b) (if $a>\fft1{\sqrt3}$).  For each of these three sets
of cases, we re-express all of the four parameters in terms of 
new variables $(x,y,z,w)$, where the ranges for each of these is
%%%%%
\bea
0\le x\le\infty\,,\quad 0\le y\le\infty\,,\quad 0\le z\le\infty,\quad
0\le w\le\infty\,.\label{xyzwranges}
\eea
%%%% 
The new parameters are introduced in such a way that the 
permitted ranges of the original $(a,\rms,q,\ell,r)$ variables translate into
the full ranges for $(x,y,z,w)$ as specified in eqn (\ref{xyzwranges}).
It is useful to note that if a quantity $p$ lies in the semi-infinite 
range $A\le p\le\infty$ (with $A>0$), then it can be reparameterised as
%%%%%
\bea
p= A\, (1+u)\,,
\eea
%%%%%
where $0\le u\le \infty$.  On the other hand, if $p$ lies in a finite
range $A\le p\le  B$, it can be reparameterised as
%%%%%
\bea
p = A + \fft{B-A}{1+u}\,,
\eea
%%%%
where again $0\le u\le\infty$.

The reparameterisations involving $y$ and $z$ are universal, in that we
take
%%%%%
\bea
\ell=2+y\,,\qquad r= \rps\, (1+z)\,.
\eea
%%%%%
(Of course $\rps$ itself is then given by eqn (\ref{solrp}).)  
The reparameterisations involving $x$ and $w$ are case-dependent, depending
upon whether we are considering cases (1), (2) and (3a), or instead 
cases (1), (2) and (3b).  For the cases specified in eqns (\ref{cases}) 
above, we then have
%%%%%
\bea
\hbox{\bf Case (1)}:&& q = 2(\ell+2)\,(1+x)\,,\qquad
       a^2= w\,,\nn\\
\hbox{\bf Case (2)}:&& q = -(\ell+2)(\ell-1) + 
\fft{\ell\,(\ell-1)}{1+x}  \,,\qquad
       a^2 = w\,,\\
\hbox{\bf Case (3a)}:&& q= -\ft43(\ell+2)(\ell-1) + 
  \fft{(1-3a^2)(\ell+2)(\ell-1)}{3(1+x)}\,,\qquad 
a^2=\fft1{3(1+w)}\,,\nn\\
\hbox{\bf Case (3b)}:&& q= -(1+a^2)(\ell+2)(\ell-1) -
            \fft{(1-3a^2)(\ell+2)(\ell-1)}{3(1+x)}\,,\quad
a^2 = \ft13 (1+w)\,.\nn
\eea
%%%%%
    Substituting these reparameterisations into the expression for 
the potential $g(a,\rms,q,\ell,r)$, we
obtain expressions that are rational functions whose numerators and 
denominators are multinomials in $(x,y,z,w)$.  (Note that 
$g(a,\rms,q,\ell,r)$ is an {\it even} function of $a$.)  After factorising
these expressions, and extracting simple overall factors that are manifestly
positive, they each boil down to a single rather complicated 
remaining multinomial 
factor $S$ in the
numerator whose positivity needs still to be analysed. Remarkably, in each of
the cases (1), (2), (3a) and (3b), it turns out that every numerical 
coefficient in the multinomial $S$ is positive.  It follows, since the
parameter ranges for $(x,y,z,w)$ are as specified in eqn (\ref{xyzwranges}),
that each of the multinomial factors $S$ is non-negative for the entire
ranges.\footnote{The number of terms in the multinomial $S$ in the
various cases are 3325, 3375, 3840 and 2674 for cases (1), (2), (3a) and
(3b) respectively.  The fact that in each case all the coefficients
turn out to be 
positive is ostensibly just a lucky circumstance, but one that is 
sufficient to imply immediately the positivity of $S$, and hence $g$.
It is far from obvious why the proof of positivity should
yield to such a convenient and conceptually simple analysis, 
since a more general multinomial could, under appropriate circumstances,
actually be positive for the 
full parameter ranges even if some of the coefficients were negative.
This would have made the proof of positivity of the multinomial
more difficult.
Similar remarks apply to our proof of positivity of the axial mode
potentials, although in that case the multinomials 
were rather less complicated.} 

   In summary, we have proven that the potentials $g_i$ in the 
wave equations for the diagonalised polar perturbations $Z^+_i$ are all
positive in the entire ranges $\rps < r < \infty$, for all allowed values of
the other parameters in the theory and black hole solutions.

\section{Conclusions}

   In this paper we have studied the linearised perturbations around the
Gibbons-Maeda static black hole solutions of the class of 
Einstein-Maxwell-Dilaton theories described by the Lagrangian (\ref{emdlag}).
These solutions generalise the Reissner-Nordstr\"om black hole
solution of Einstein-Maxwell theory, and our methods for studying the
perturbations were parallel to those developed by Chandrasekhar and 
Xanthopoulos for studying the perturbations of Reissner-Nordstr\"om.

  The Gibbons-Maeda black holes are static and spherically
symmetric, and one can without loss of generality restrict attention to
perturbations that are independent of the azimuthal angle $\varphi$.  A further
simplification arises because the black hole background is parity-invariant.
This has the consequence that the linearised modes fall into two 
disjoint sets, namely the axial modes, which are of odd parity under
sending $\varphi\rightarrow -\varphi$, and the polar modes, which are of
even parity.  It was already the case in the study of the 
Reissner-Nordstr\"om perturbations by Chandrasekhar and Xanthopoulos that
the analysis of the modes in the polar sector was much more complicated 
than the analysis of the axial modes. This disparity arose because
after separating the fluctuation equations in the
polar sector, the radial equations in the Reissner-Nordstr\"om case
were described by a reducible system of five first-order equations.
A rather intricate analysis of this system was necessary in order
to show that the general solution could be written in terms of the
solutions to two decoupled second-order equations. By contrast, in
the axial sector it was relatively straightforward to derive directly
two decoupled second-order equations whose solutions characterised the
general axial perturbations.

   In the case of the perturbations around the Gibbons-Maeda solutions of
Einstein-Maxwell-Dilaton theories that we studied in this paper, the disparity
between the axial and the polar sectors is rather greater.  This is because 
the perturbations of the additional dilaton field of EMD theory are polar,
leading to an enlarged set of polar equations that are described by 
a reducible system of seven first-order equations. Nevertheless, we were
able to analyse this system completely, and to show that the general
solution for the polar perturbations is characterised by the general solutions
to three decoupled second-order equations.  By contrast, the axial
sector is again more straightforward to analyse, and the perturbations
here are characterised by the general solutions to two decoupled second-order
equations.  

  The fact that the perturbations of the Gibbons-Maeda black holes
are characterised by two solutions of second-order equations in the axial
sector, and by three solutions of second-order equations in the polar sector,
highlights a significant difference from the situation in the 
Reissner-Nordstr\"om case.  For Reissner-Nordstr\"om, where there
are two second-order equations in the axial sector and two second-order
equations in the polar sector, Chandrasekhar and
Xanthopoulos were able to show that there existed fairly simple mappings
between the potentials and eigenfunctions of the axial sector and those
of the polar sector. In particular, the potentials $V^-_i$ and $V^+_i$
are isospectral in the Reissner-Nordstr\"om case. 

On the face of it, no such relations seem likely to
arise for the perturbations of the Gibbons-Maeda black holes in the EMD
theories.  First of all, there is now a mismatch between the
number of second-order equations in the axial and the polar sectors.  
In addition, the natures of the potentials in the two sectors are
now very different, being characterised by solutions of a quadratic
equation in the axial sector but a cubic equation in the polar sector.

This raises an intriguing further question because in the Reissner-Nordstr\"om
case, the relation between the eigenfunctions and potentials in the axial
and the polar sectors can be understood from the fact that the two kinds
of perturbations are treated in a uniform manner in an alternative
approach to studying the problem by using the Teukolsky description of
gauge-invariant curvature perturbations.  It will be interesting, therefore,
to study the perturbations of the Gibbons-Maeda black holes within the
framework of the Teukolsky approach.  This is currently under investigation.

A generalisation of the work in this paper would be to consider dyonic black
hole solutions in the Einstein-Maxwell-Dilaton theories.  These are known
explicitly for special values of the dilaton coupling $a$, namely $a=0$, $a=1$
and $a=\sqrt3$, but solutions will exist for all values of $a$.  One of the
complications that will arise for the dyonic black holes is that the background
solutions are no longer parity-invariant, meaning that the decoupling of the
perturbations in the axial and polar sectors will no longer occur.  This 
substantially complicates the analysis.  We intend to report on this in
forthcoming work.

Another focus of our investigations is to consider the black hole solutions
in supergravity theories and more generally in theories with larger numbers of
scalar fields and, possibly, Maxwell and higher-form fields (in more than four
dimensions).

\section*{Acknowledgements}

We are grateful to Mihalis Dafermos and Yakov Shlapentokh-Rothman for
helpful discussions.
C.N.P.~and D.O.R.~are supported in part by DOE grant DE-SC0010813.  
B.F.W.~is supported in part by NSF grant PHY 1607323.

\appendix
\section{The Matrix $\bM$}

  Here, we present a detailed construction of the matrix $\bM$ that appears in
the second-order equation (\ref{secondorderop}). As was seen in section 
\ref{sec:polarmodes}, the eigenfunctions $Z^+$ of the diagonalised second-order 
equation (\ref{Zeqn}) take the general form given in eqn (\ref{Zans}), 
with the various functions $(f_1,f_2,f_3,f_4,f_5)$ then taking the
forms found later in the section.  Eventually, it was found that, up to
irrelevant overall constant scaling, there are three independent 
eigenfunctions that satisfy eqn (\ref{Zeqn}), corresponding to the 
constants $c_1$ and $c_4$ being given in terms of $q$ by eqns 
(\ref{c1q}) and (\ref{c4q}), and then the dimensionless constant
$q$ obeying the cubic equation (\ref{cubic}), 
hence leading to the three independent eigenfunctions. 

  The three independent eigenfunctions of eqn (\ref{Zeqn}) correspond to
choosing the three roots $q_1$, $q_2$ and $q_3$ of the cubic (\ref{cubic}),
and then plugging the expressions (\ref{c1q}) and (\ref{c4q}) for 
$c_1$ and $c_4$ into the results for $(f_1,f_2,f_3,f_4,f_5)$ found in
section \ref{sec:polarmodes}. (The constant $c_3$ will now be an unimportant
overall scale factor in the eigenfunctions, and it will cancel out
completely in the expression for $g$ in eqn (\ref{Zeqn}).)  We
denote the three eigenfunctions by $Z^+_i$, with corresponding potential
functions $g_i$ appearing on the right-hand side of eqn (\ref{Zeqn}).  Thus
we shall have
%%%%%
\bea
(\del_{r_*}^2 + \omega^2)\, Z^+_i = g_i\, Z^+_i\,,\qquad \hbox{for}\quad
i=1, 2, 3\,.\label{Zieqn}
\eea
%%%%%

    We may now relate these eigenfunctions $Z^+_i$ to the functions 
$(H^+_1,H^+_2,H^+_3)$ that we introduced in section \ref{sec:genpolar}.  In
view of eqn (\ref{ZH0}) and the rescalings 
specified by (\ref{Hrescal}) and (\ref{kchoice}), we have
%%%%%
\bea
Z^+= -Q\,\mu\, c_3 \, H^+_1 + n\, c_1\, H^+_2 + 2\sqrt{n(n+1)}\, c_4\, H^+_3\,,
\eea
%%%%%
where as usual $\mu=\sqrt{2n}$.  After the reparameterisation (\ref{c1q})
and using eqn (\ref{c4q}), we shall then have
%%%%%
\bea
Z^+_i = -Q\,\mu\, H^+_1 + \fft{q_i\,\rms}{2(1+a^2)}\, H^+_2 + 
  \fft{2a\,q_i\,\rms\, \sqrt{n(n+1)}}{2(1+a^2)\, (q_i+2n)}\, H^+_3\,,
\eea
%%%%%
corresponding to the three roots $q_i$ of the cubic equation 
(\ref{cubic}).\footnote{We have without loss of generality set
the constant $c_3$, which is now just an overall scale factor, to unity.}
Substituting these expressions into eqn (\ref{Zieqn}) it is then 
a straightforward matter to take the appropriate linear combinations in
order to arrive equations of the form (\ref{secondorderop}), and hence
to read off the various components $M_{ij}$ of the matrix $\bM$.  Doing this,
we find
%%%%%
\bea
M_{11}&=& \fft{g_1\, q_2\,q_3\,(q_1+2n)}{2n\,(q_1-q_2)(q_1-q_3)} 
   +\hbox{cyclic}\,,\nn\\
M_{12} &=& M_{21}= -\fft{g_1\, q_1\, q_2\, q_3\, (q_1+2n\, \rms)}{
  4\sqrt{2}\, (1+a^2)\, n^{3/2}\, Q\, (q_1-q_2)(q_1-q_3)}+
       \hbox{cyclic}\,,\nn\\ 
M_{13}&=& M_{31}= -\fft{a\, g_1\, q_1\, q_2\, q_3\,\rms\, \sqrt{n+1}}{
        2\sqrt2\, n\, (1+a^2)\, Q\, (q_1-q_2)(q_1-q_3)} 
       +\hbox{cyclic}\,,\nn\\
M_{22} &=& \fft{g_1\, q_1\, (q_1+2n)}{(q_1-q_2)(q_1-q_3)}
   +\hbox{cyclic}\,,\nn\\
M_{23}&=& M_{32} = \fft{2a\, g_1\, q_1\, \sqrt{n(n+1)}}{
(q_1-q_2)(q_1-q_3)} +\hbox{cyclic}\,,\nn\\
M_{33}&=& -\fft{g_1\, q_1\, (q_2+2n)\,(q_3 + 2n)}{
  2n\, (q_1-q_2)(q_1-q_3)}
   +\hbox{cyclic}\,.\label{Mij}
\eea
%%%%%
In each case the notation ``$+\,\hbox{cyclic}$'' 
indicates that two further terms
are to be added, corresponding to the two further cyclings of the subscripts
1, 2 and 3.  For example, the full expression for $M_{11}$ is
%%%%%
\bea
M_{11}&=& \fft{g_1\, q_2\,q_3\,(q_1+2n)}{2n\,(q_1-q_2)(q_1-q_3)}
  + \fft{g_2\, q_1\,q_3\,(q_2+2n)}{2n\,(q_2-q_1)(q_2-q_3)} +
 \fft{g_3\, q_1\,q_2\,(q_3+2n)}{2n\,(q_3-q_1)(q_3-q_2)}\,,\nn\\
&=& \sum_i \fft{q_1\, q_2\, q_3\, (q_i+2n)}{2n\, q_i\,
\prod_{j\ne i}(q_i-q_j)}\, g_i\,.
\eea
%%%%%

  The functions $g_i$ are given by eqn (\ref{gres}), with the constants
$c_1$ and $c_4$ given as in eqns (\ref{c1q}) and (\ref{c4q}), and with $q$
then being the root $q_i$ of the cubic equation (\ref{cubic}).  Note that
because of the cyclic symmetry of all the expressions for the components
$M_{ij}$ in eqns (\ref{Mij}), it follows that the roots $q_i$ enter only via
the symmetric combinations $(q_1+q_2+q_3)$, \ $(q_1\, q_2 + q_2\, q_3+
q_1\, q_3)$ and $q_1\, q_2\, q_3$, and therefore the components 
of the matrix $\bM$ are expressible solely in terms of $(r,\rms,\rps,a,n)$. 
The explicit expressions are easily obtainable by computer, but they are
rather too complicated to be worth presenting here explicitly.

\subsection{A general lemma}

Looking at the function $g$ in eqn (\ref{gres}), it can be seen
that after replacing the $c_i$ coefficient in terms of $q$ as described
above, we will have
%%%%%
\bea
g_i = x_1 + x_2\, q_i + \fft{x_3}{q_i} \,,
\eea
%%%%%
where the functions $x_i$, which are rather complicated rational 
functions of $r$, 
are independent of the $q$'s, and can be read off 
from the expression for $g$ in eqn (\ref{gres}). Substituting
this into the expression, for example, for $M_{11}$ in eqns (\ref{Mij}),
one can easily then see that
%%%%
\bea
M_{11}= x_1 - \fft{3(1+a^2)\, \rps + (3-a^2)\, \rms}{8n\, (1+a^2)\, \rps}\, 
  x_3\,,
\eea
%%%%%
after using the cubic equation (\ref{cubic}) to replace the
symmetric combinations of the $q_i$ according to
%%%%
\bea
q_1\, q_2\, q_3 &=& \fft{16 n^2\, (1+a^2)\,\rps}{\rms}\,,\nn\\
q_1\, q_2 + q_1\, q_3 + q_2\, q_3 &=& -\fft{2n\,\big[(3-a^2)\,\rms
                  + 7(1+a^2)\, \rps\big]}{\rms}\,,\nn\\
q_1+q_2+q_3 &=& \fft{3(1-a^2)\, \rms +3(1+a^2)\, \rps -2n\,(1+a^2)\,\rms}{
             \rms} \,.
\eea
%%%%%
Similar calculations can be performed to obtain expressions for all the
other components of $\bM$.

\section{Specialisation To Reissner-Nordstr\"om}
 
   In the limit where the dilaton coupling $a$ is sent to zero, the
Einstein-Maxwell-Dilaton system reduces to Einstein-Maxwell together
with a massless scalar field that couples only to gravity.  The static EMD
black holes that have formed the background in our analysis reduce to
Reissner-Nordstr\"om black holes, with the dilaton being zero in the 
background.  At the level of linearised perturbations, the coupled system
of gravitational, electromagnetic and dilaton fluctuations that
we studied in this paper reduces to the coupled gravitational and
electromagnetic fluctuations around the Reissner-Nordstr\"om background
that were studied in \cite{xanth2,chandra}, together with a decoupled
massless scalar fluctuation that obeys the massless Klein-Gordon
equation in the Reissner-Nordstr\"om background geometry.

    One can see explicitly that our results in this paper 
indeed reduce to those in \cite{xanth2,chandra} for
Reissner-Nordstr\"om, if
we set the dilaton coupling $a=0$.  We then find that the cubic equation
(\ref{cubic}) reduces to
%%%%%
\bea
(q+2n)(\rms \, q^2 -3(\rms+\rps)\, q
  -8 n\, \rps)=0\,.\label{cubictoquad}
\eea
%%%%%
Defining $q_{\sst{RN}}= \rms\, q$ and using $\mu=\sqrt{2n}$,
the quadratic factor gives the quadratic $(q_{\sst{RN}}^2 -6M\, q_{\sst{RN}}
-4\mu^2\, Q^2)$, where $M=\ft12 (\rms+\rps)$ and $Q=\sqrt{\rms\,\rps}$, with
the roots
%%%%%
\bea
q_{\sst{RN}} = 3M \pm \sqrt{9M^2+4Q^2\,\mu^2}\,,
\eea
%%%%%
which agrees with the  expressions for $q_1$ and $q_2$ in eqn
(149) of section 42 of \cite{chandra}.  (The linear factor in eqn
(\ref{cubictoquad}) represents the dilaton mode, which decouples in
the $a\rightarrow 0$ Reissner-Nordstr\"om limit.)

The function $\Xi$ defined in eqn (\ref{defXi}) reduces to
%%%%%
\bea
\Xi &\longrightarrow& (r-\rms)
             (2n\, r^2 + 3(\rms+\rps)\, r - 4\rms\, \rps)\nn\\
&=& 2 r\, (r-\rms)\Big(n\, r + 3M - \fft{2Q^2}{r}\Big)\,.
\eea
%%%%%
The final factor here is precisely the same as the function
$\varpi$ defined in \cite{chanxant,xanth2}.

%\coffeestainA{0.9}{0.85}{-25}{5cm}{1.3cm}

\end{document}